\documentclass[aps,prd,twocolumn,nofootinbib,floatfix]{revtex4} 
\pdfoutput=1

\usepackage{graphicx}

\usepackage{amsmath}
\usepackage{amssymb}

\font\FermiSmallfont=cmssq8 scaled 1200

\def\UMDppthead#1#2#3{
\null 
\begin{center}\vskip -1.0truein{\hbox to 7.5truein {
\hfill
\vbox to 1in {\vfill \FermiSmallfont
              \hbox{#1}
              \hbox{#2}
              \hbox{#3}
              \vfill}
}}\vskip-0.0truein\end{center}}%FNALppthead

\begin{document}

\UMDppthead{UMD-PP-10-002}{RUNHETC-2010-03}{arXiv:1002.3820v3}

\title{Conservative Constraints on Dark Matter from
  the Fermi-LAT Isotropic Diffuse Gamma-Ray Background Spectrum}

 % \author{Kevork N.\ Abazajian\\
%  Maryland Center for Fundamental Physics, Department of
%    Physics, University of Maryland, College Park, Maryland  20742  USA}

%  \author{Prateek Agrawal\\ Maryland Center for Fundamental Physics, Department of
%    Physics, University of Maryland, College Park, Maryland  20742  USA}

%  \author{Zackaria Chacko\\
%   Maryland Center for Fundamental Physics, Department of
%    Physics, University of Maryland, College Park, Maryland  20742  USA}

%   \author{Can Kilic\\ Department of Physics \& Astronomy,
%     Rutgers University, Piscataway, NJ 08854, USA}

 \author{Kevork N.\ Abazajian}
 \affiliation{Maryland Center for Fundamental Physics, Department of
   Physics, University of Maryland, College Park, Maryland  20742  USA}

 \author{Prateek Agrawal}
 \affiliation{Maryland Center for Fundamental Physics, Department of
   Physics, University of Maryland, College Park, Maryland  20742  USA}

 \author{Zackaria Chacko}
 \affiliation{Maryland Center for Fundamental Physics, Department of
   Physics, University of Maryland, College Park, Maryland  20742  USA}

 \author{Can Kilic} \affiliation{Department of Physics \& Astronomy,
   Rutgers University, Piscataway, NJ 08854, USA}

% \pacs{95.35.+d, 95.85.Pw, 98.70.Vc, 98.62.Gq}  

%\abstract{
\begin{abstract} 
 We examine the constraints on final state radiation from Weakly
  Interacting Massive Particle (WIMP) dark matter candidates
  annihilating into various standard model final states, as imposed by
  the measurement of the isotropic diffuse gamma-ray background by the
  Large Area Telescope aboard the Fermi Gamma-Ray Space Telescope.
  The expected isotropic diffuse signal from dark matter annihilation
  has contributions from the local Milky Way (MW) as well as from
  extragalactic dark matter.  The signal from the MW is very
  insensitive to the adopted dark matter profile of the halos, and
  dominates the signal from extragalactic halos, which is sensitive to
  the low mass cut-off of the halo mass function. We adopt a
  conservative model for both the low halo mass survival cut-off and
  the substructure boost factor of the Galactic and extragalactic
  components, and only consider the primary final state radiation.
  This provides robust constraints which reach the thermal production
  cross-section for low mass WIMPs annihilating into hadronic modes.
  We also reanalyze limits from HESS observations of the Galactic
  Ridge region using a conservative model for the dark matter halo
  profile.  When combined with the HESS constraint, the isotropic
  diffuse spectrum rules out all interpretations of the PAMELA
  positron excess based on dark matter annihilation into two lepton
  final states. Annihilation into four leptons through new
  intermediate states, although constrained by the data, is not
  excluded.
%}
\end{abstract}

%\begin{document}

\maketitle

\section{Introduction}
The identity of the dark matter has remained a fundamental unsolved
problem in cosmology and particle physics for nearly eighty
years~\cite{Zwicky:1933gu}.  The likelihood of new physics near the
electroweak scale suggests a particle dark matter candidate of mass
$\sim$100 GeV.  A weak-scale interaction strength for such a particle
can naturally produce a relic abundance at the observed dark matter
density.  Such a weakly-interacting massive particle (WIMP) arises in
supersymmetric extensions of the standard model as well as other
beyond the standard model extensions.  For a review of particle dark
matter candidates, see, e.g.,
Refs.~\cite{Jungman:1995df,Bertone:2004pz}.

Thermal production of dark matter prefers a scale of the dark matter
annihilation cross-section at $\langle \sigma_{\rm A} v\rangle \approx
3\times 10^{-26}\rm\ cm^3\ s^{-1}$.  This scale predicts a potentially
detectable gamma-ray photon signal from annihilations in local and
cosmological dark matter structures, {\it e.g.}, see
Ref.~\cite{Springel:2008by}. Annihilation channels directly to photon
pairs produce a distinctive $\gamma$-ray spectral line, while channels
to charged standard model particles produce an associated continuum
emission of $\gamma$-rays from bremsstrahlung radiation and in
hadronization via $\pi^0\rightarrow \gamma\gamma$.  The Large Area
Telescope (LAT) aboard the recently launched Fermi Gamma-Ray Space
Telescope has significant sensitivity to such $\gamma$-ray radiation
from dark matter annihilation.  The sensitivity of Fermi-LAT to a an
annihilation signal from the Galactic center (GC), Galactic satellites
and the isotropic diffuse signal was reviewed in Baltz et
al.~\cite{Baltz:2008wd}.

Recent Fermi-LAT observations have led to constraints on dark
matter annihilation, including observations of dwarf galaxies analyzed
by the Fermi-LAT
Collaboration~\cite{Scott:2009jn,Abdo:2010ex}, as well as
analyses of public data of the total sky flux~\cite{Papucci:2009gd},
and public data from selected portions of the
sky~\cite{Cirelli:2009dv}.  Most recently, the Fermi-LAT
collaboration released the spectrum from the isotropic diffuse
background in $\gamma$-rays~\cite{Abdo:2010nz}.  Here, we use
the Fermi-LAT Collaboration derived diffuse-photon spectrum in this
work to place constraints on dark matter annihilation channels.
Depending on the background estimation, the relative sensitivity of
these methods varies.

In this work, we constrain the partial cross section for dark matter
annihilation to various particle anti-particle standard model 
final states using the diffuse
isotropic $\gamma$-ray photon spectrum from the
Fermi-LAT~\cite{Abdo:2010nz}.
The Fermi-LAT isotropic diffuse
spectrum uses enhanced event simulation modeling that improves the
background rejection in the observations by a factor of 1.3-10.
Therefore, we expect the spectrum to be a stringent constraint on dark
matter annihilation.  The errors on the estimate of the diffuse
isotropic background come from systematic errors in modeling
foreground Galactic diffuse $\gamma$-ray emission.  The isotropic
diffuse background could include a dark matter annihilation signal,
yet also has contributions from unresolved active galactic nuclei,
starburst galaxies and $\gamma$-ray bursts, as well as any truly
diffuse components in large scale structure.  In total, the modeling
of the Galactic diffuse emission, as well as the extragalactic
$\gamma$-ray emission processes, are the ultimate observational limits
to the sensitivity of the isotropic diffuse background to a dark
matter signal.  

Since Cold Dark Matter (CDM) particle candidates such as the WIMPs
considered here produce cusped
halos~\cite{Navarro:1996gj,Springel:2008by}, in applying our
constraints, we use conservative models for the expected dark matter
profile density distribution which are consistent with the cusped
profiles produced by such CDM candidates.  We also apply this
requirement to update limits from existing constraints from
observations of the GC ridge by HESS.

The Fermi-LAT collaboration performed a similar analysis of the
constraints from the diffuse background
observations~\cite{Abdo:2010dk}. That work did not include the minimal
Galactic contribution that is necessarily present in the observation,
and is therefore less stringent than the limits presented here.
Previous work along these lines prior to Fermi-LAT data include limits
from older $\gamma$-ray observations with a subset of standard model
channels: Bell \& Jacques~\cite{Bell:2008vx} examined constraints on
the branching ratio to $e^+/e^-$ with the EGRET, CGRO, HESS and
CELESTE data, and Mack et al.~\cite{Mack:2008wu} applied constraints
on lines in the case of the two-photon channel with a similar set of
data.  With the dawn of the Fermi-LAT data, constraints from all-sky
and diffuse flux limits have been made with public data.  Cirelli et
al.~\cite{Cirelli:2009dv} use the total flux from specific sky regions
to constrain dark matter annihilations using conservative models for
the dark matter halo profile. They include $\gamma$-rays arising from
inverse-Compton scattering (ICS) of annihilation products on cosmic
microwave background and inter-stellar photons.  In a similar
analysis, Pappucci \& Strumia~\cite{Papucci:2009gd} find limits to
dark matter annihilation modes from a full-sky analysis of the public
Fermi-LAT data, including final state radiation limits as well as ICS,
with no reduction of the sky flux due to foreground galaxy and point
source removal.  Therefore, their constraints are not as stringent as
ours or those in Ref.~\cite{Cirelli:2009dv}.
Ref.~\cite{Papucci:2009gd} also overstates MW halo profile
dependencies by adopting isothermal halo profile models, which, as we
discuss below, are inconsistent with the CDM structure formation
paradigm for a WIMP candidate.

There has been considerable interest in the possibility of dark matter
annihilation as being the source of the excess cosmic ray positron
fraction at $\sim$10-100~GeV observed by HEAT\footnote{High Energy
  Antimatter Telescope balloon experiment}~\cite{Barwick:1997ig},
AMS-01\footnote{The Alpha Magnetic Spectrometer
  experiment}~\cite{Aguilar:2007yf} and PAMELA\footnote{Payload for
  Antimatter Matter Exploration and Light-nuclei Astrophysics,
  http://pamela.roma2.infn.it}~\cite{Adriani:2008zr}, where $e^+/e^-$
pairs are produced directly or indirectly in a dark matter particle
pair annihilation cascade~\cite{Baltz:2001ir,Cholis:2008hb}.  In
addition, features in the higher-energy 10$^2$-10$^3$~GeV positron
spectrum seen by ATIC\footnote{Advanced Thin Ionization Calorimeter,
  http://atic.phys.lsu.edu/aticweb/}~\cite{Chang:2008zzr} and
Fermi-LAT~\cite{Abdo:2009zk} are also consistent with the dark matter
annihilation interpretations of in the lower energy positron excess
data~\cite{Cirelli:2008pk,Bergstrom:2009fa,Meade:2009iu}.  In order to
achieve the dark matter annihilation rate required for these $e^+/e^-$
signals that may be consistent with the expected thermal production
cross section, and to avoid an excess in anti-proton observations, the
annihilation rate can be enhanced through a low-energy Sommerfeld
enhancement, and limited to leptonic modes with a $<$1 GeV dark-force
carrying
particle~\cite{Cholis:2008vb,ArkaniHamed:2008qn,Baumgart:2009tn,Katz:2009qq}.
Such an enhanced cross-section from a new force is in tension with
detailed calculations of the relic abundance of the dark matter, so
that such a candidate may not contribute to all of the dark
matter~\cite{Zavala:2009mi,Feng:2009hw,Buckley:2009in}, and are also
constrained by nonthermal distortions of the CMB~\cite{Zavala:2009mi}
and asphericity observed in dark matter halos~\cite{Feng:2009hw}.

These models are also constrained by corresponding Fermi-LAT
$\gamma$-ray observations of dwarf galaxies \cite{Abdo:2010ex}, the
total sky flux \cite{Papucci:2009gd}, portions of the sky
\cite{Cirelli:2009dv}, Galactic radio synchrotron
emission~\cite{Bertone:2008xr}, the neutrino flux from the GC as
observed by SuperKAMIOKANDE~\cite{Meade:2009iu}, and atmospheric
Cerenkov observations of $\gamma$-rays from dwarf
galaxies~\cite{Essig:2009jx}.  H\"utsi et al.~\cite{Huetsi:2009ex}
have used diffuse $\gamma$-ray observations by EGRET to constrain dark
matter models that can explain the PAMELA data, but as we explain
below, with more optimistic assumptions for the low mass halo cutoff
and extragalactic signal.  The PAMELA, HEAT, and Fermi $e^+/e^-$
signals are also consistent with high-energy $e^+/e^-$ emission from
pulsars and supernova
remnants~\cite{Boulares1989,Hooper:2008kg,Yuksel:2008rf,Profumo:2008ms}.
Here, we show how the dark matter annihilation interpretation of
theses signals is in conflict with the observed isotropic diffuse flux
spectrum of Fermi-LAT in combination with other constraints for
two-body standard model particle final states, and constrains scalar
or vector boson mediated four-lepton final states.

%\FIGURE{
\begin{figure}
\includegraphics[width=3.2truein]{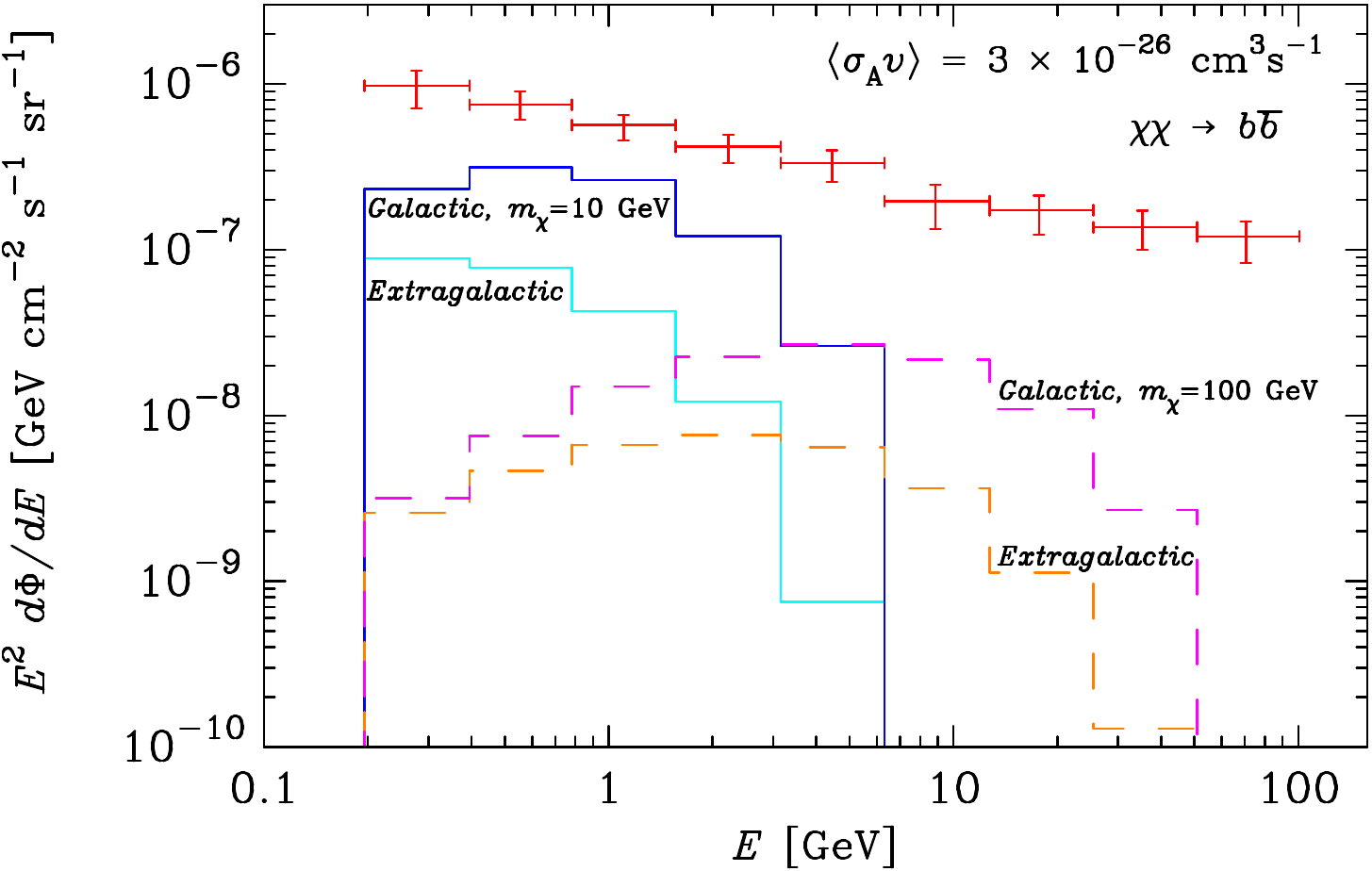}
\caption {\small Shown is the observed isotropic diffuse
  background spectrum observed by Fermi-LAT (points with errors), and
  representative models of the annihilation spectrum of the Galactic
  (upper) and extragalactic (lower) contributions
  from the channel $\chi\chi\rightarrow b\bar{b}$ for $m_\chi = 10\rm\
  GeV$ (solid), and for $m_\chi = 100\rm\ GeV$ (dashed).
  Both cases have the annihilation cross section $\langle \sigma_{\rm A}
  v\rangle = 3\times 10^{-26}\ \rm cm^3\ s^{-1}$.
\label{flux_plot}}
%}
\end{figure}

\begin{figure*}
%\FIGURE{
% \includegraphics[width=1.95truein]{ee_0000dat.pdf}
% \includegraphics[width=1.95truein]{uubar_0dat.pdf}
% \includegraphics[width=1.95truein]{ddbar_0dat.pdf}\\
% \includegraphics[width=1.95truein]{mumu_00dat.pdf}
% \includegraphics[width=1.95truein]{ccbar_0dat.pdf}
% \includegraphics[width=1.95truein]{ssbar_0dat.pdf}\\
% \includegraphics[width=1.95truein]{tautau_dat.pdf}
% \includegraphics[width=1.95truein]{ttbar_0dat.pdf}
% \includegraphics[width=1.95truein]{bbbar_0dat.pdf}\\
% \includegraphics[width=1.95truein]{hh120_0dat.pdf}
% \includegraphics[width=1.95truein]{ZZ_0000dat.pdf}
% \includegraphics[width=1.95truein]{WW_0000dat.pdf}
 \includegraphics[width=2.2truein]{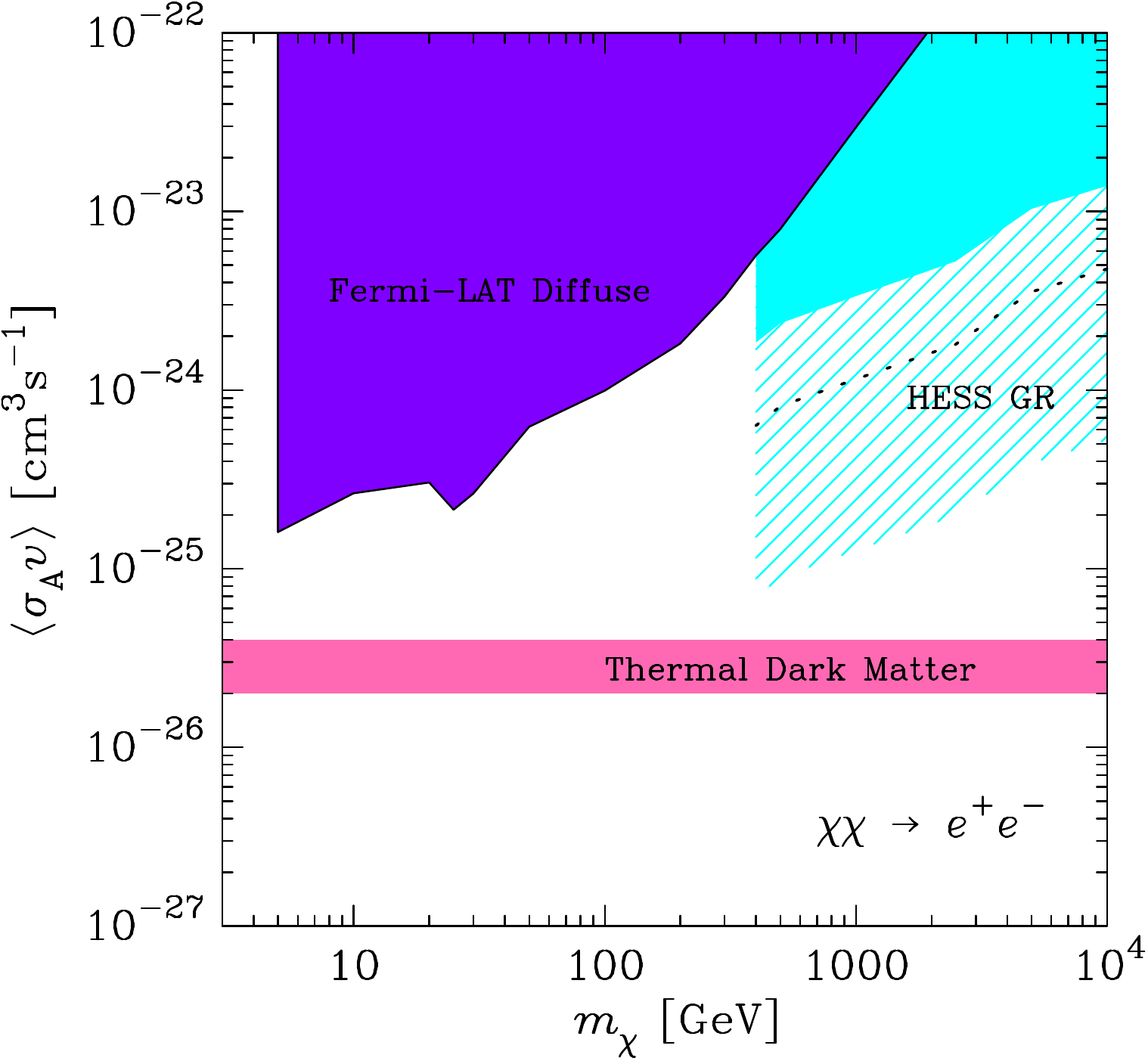}
 \includegraphics[width=2.2truein]{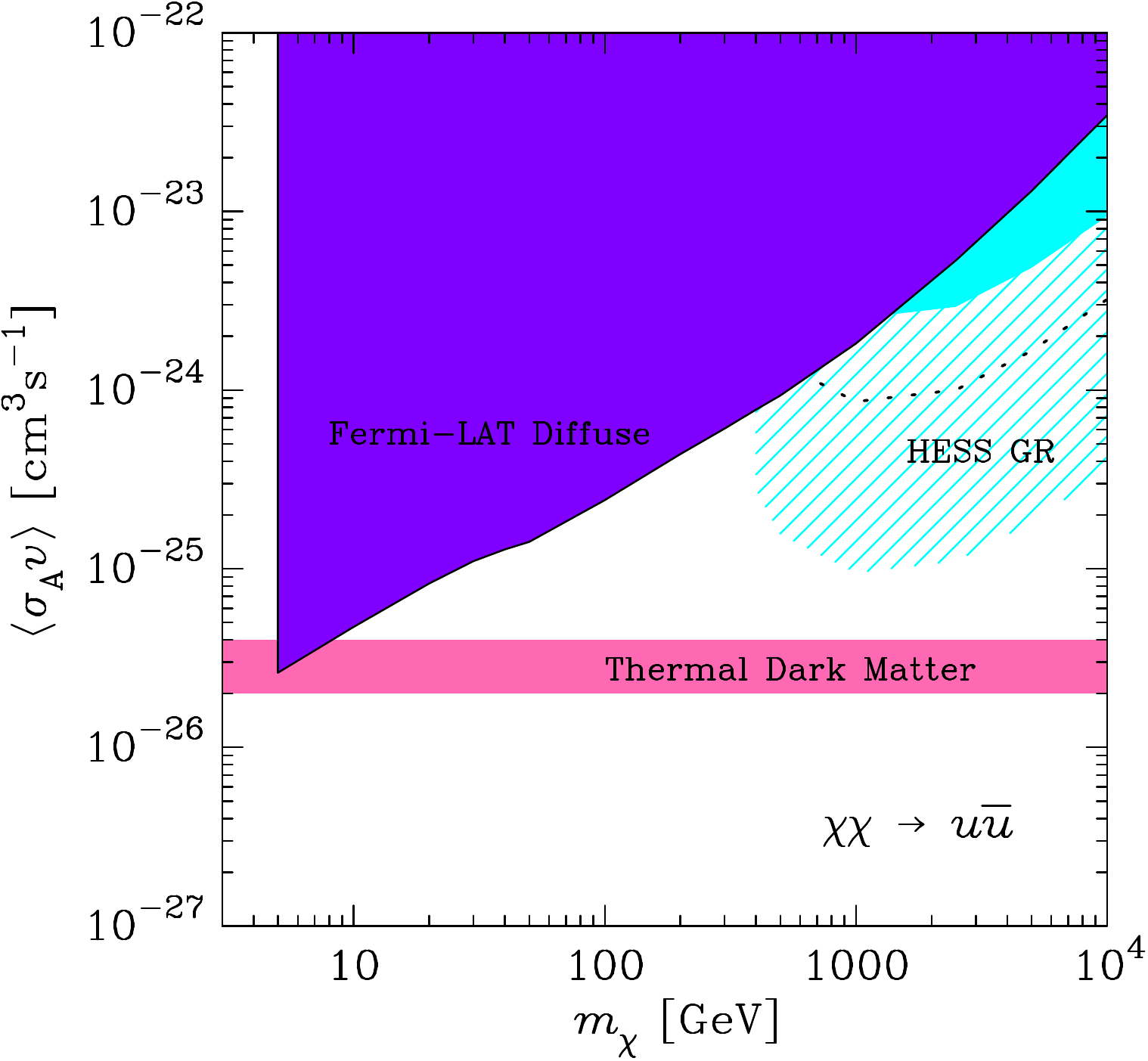}
 \includegraphics[width=2.2truein]{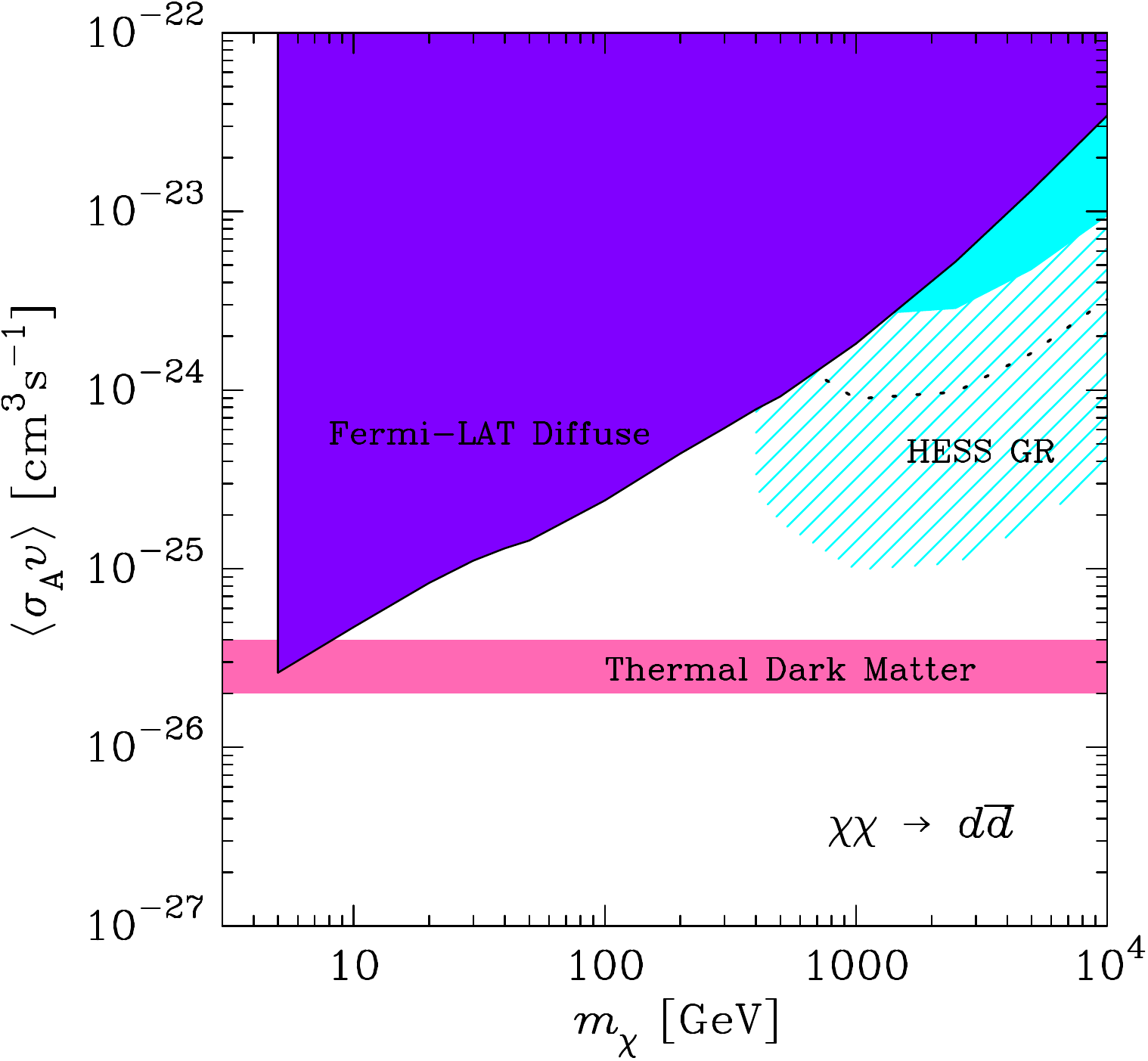}\\
 \includegraphics[width=2.2truein]{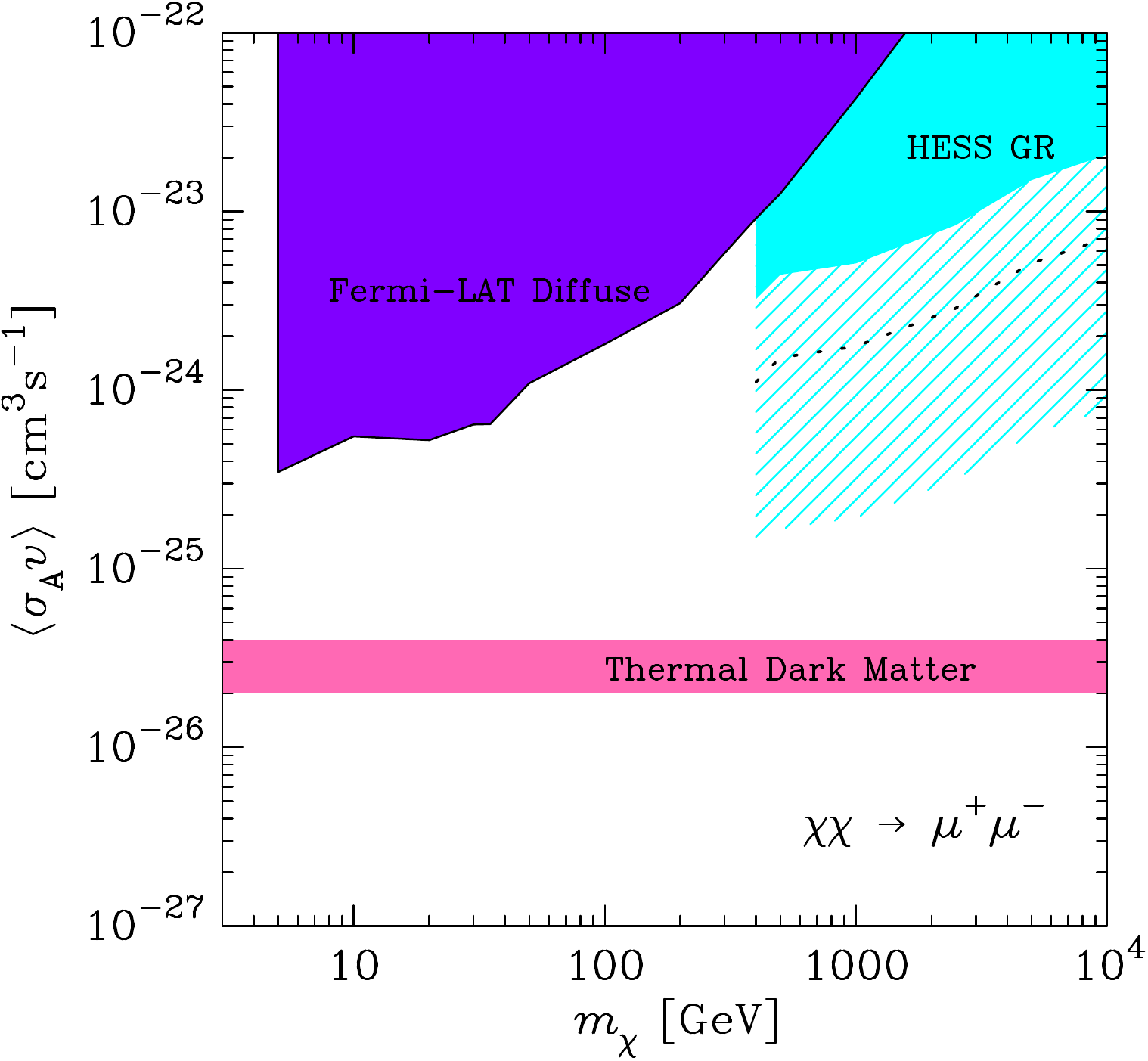}
 \includegraphics[width=2.2truein]{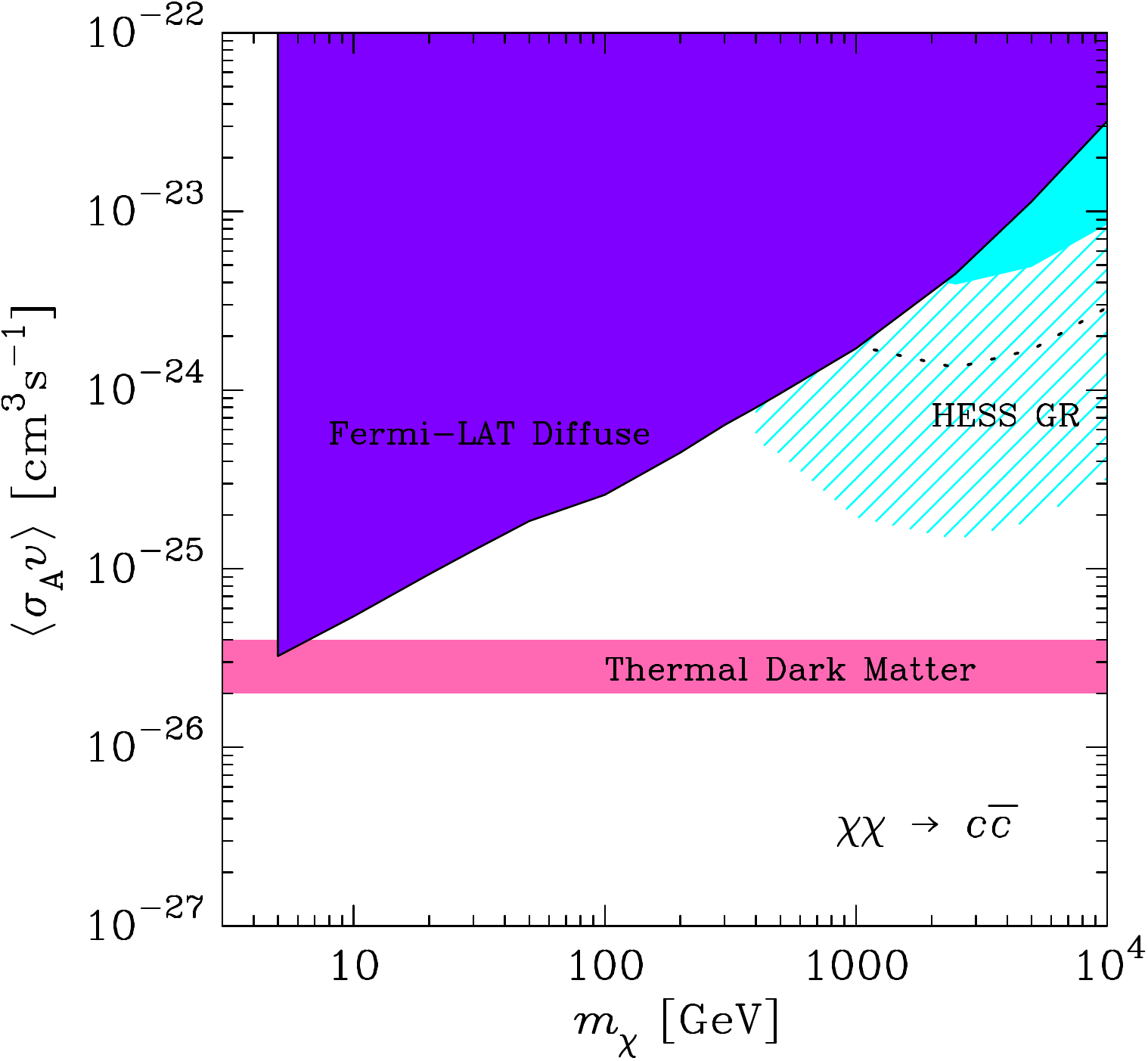}
 \includegraphics[width=2.2truein]{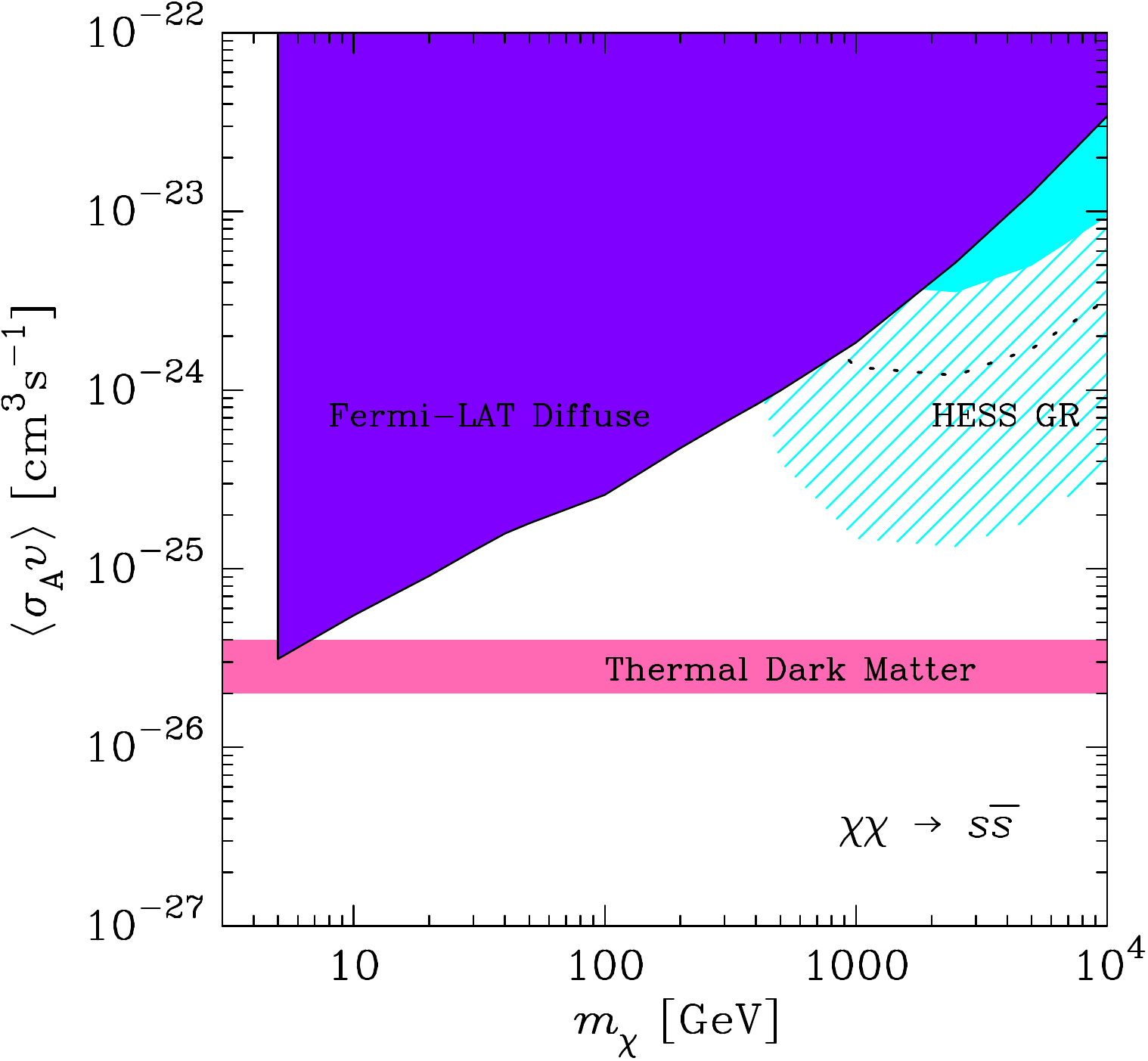}\\
 \includegraphics[width=2.2truein]{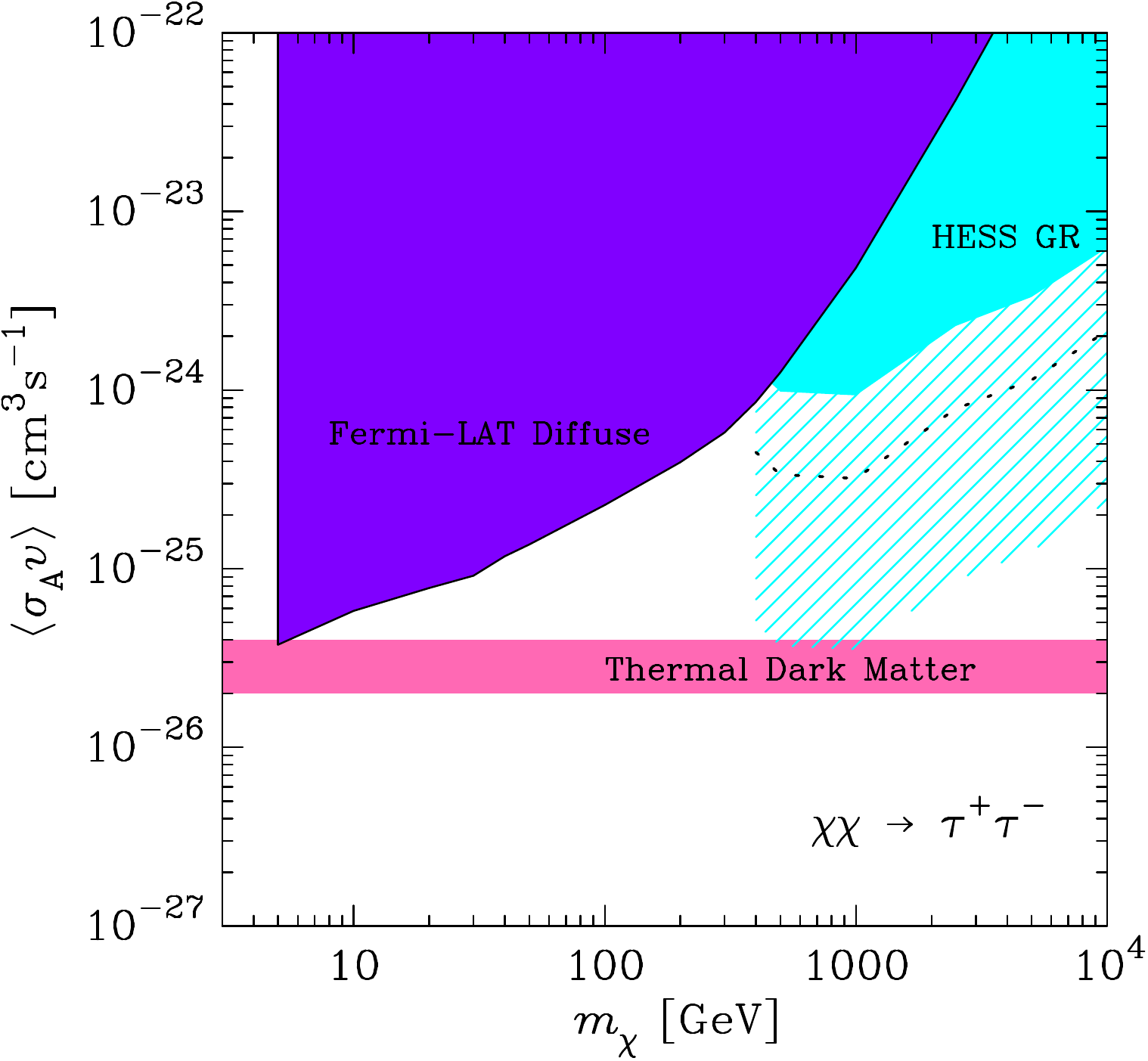}
 \includegraphics[width=2.2truein]{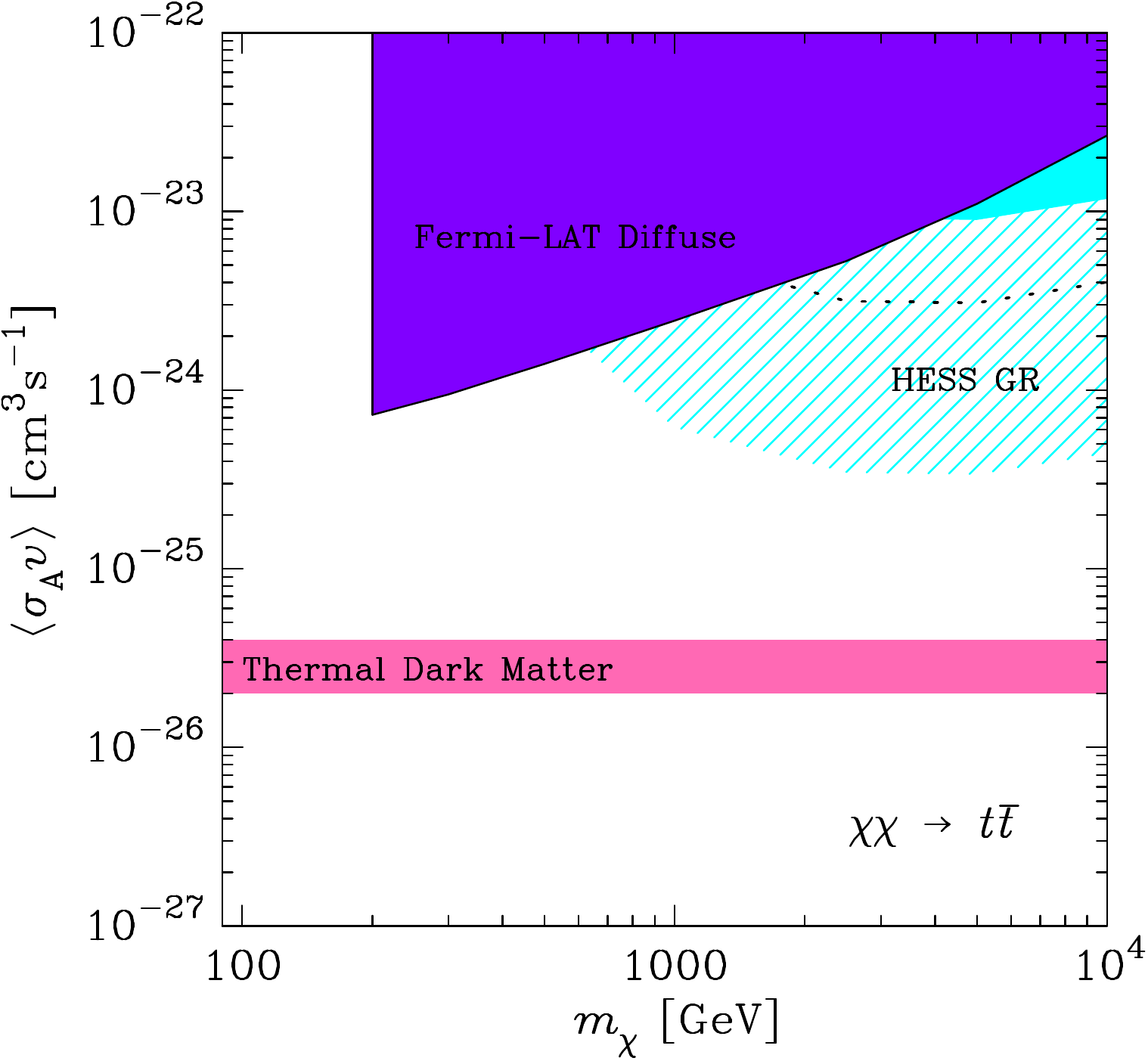}
 \includegraphics[width=2.2truein]{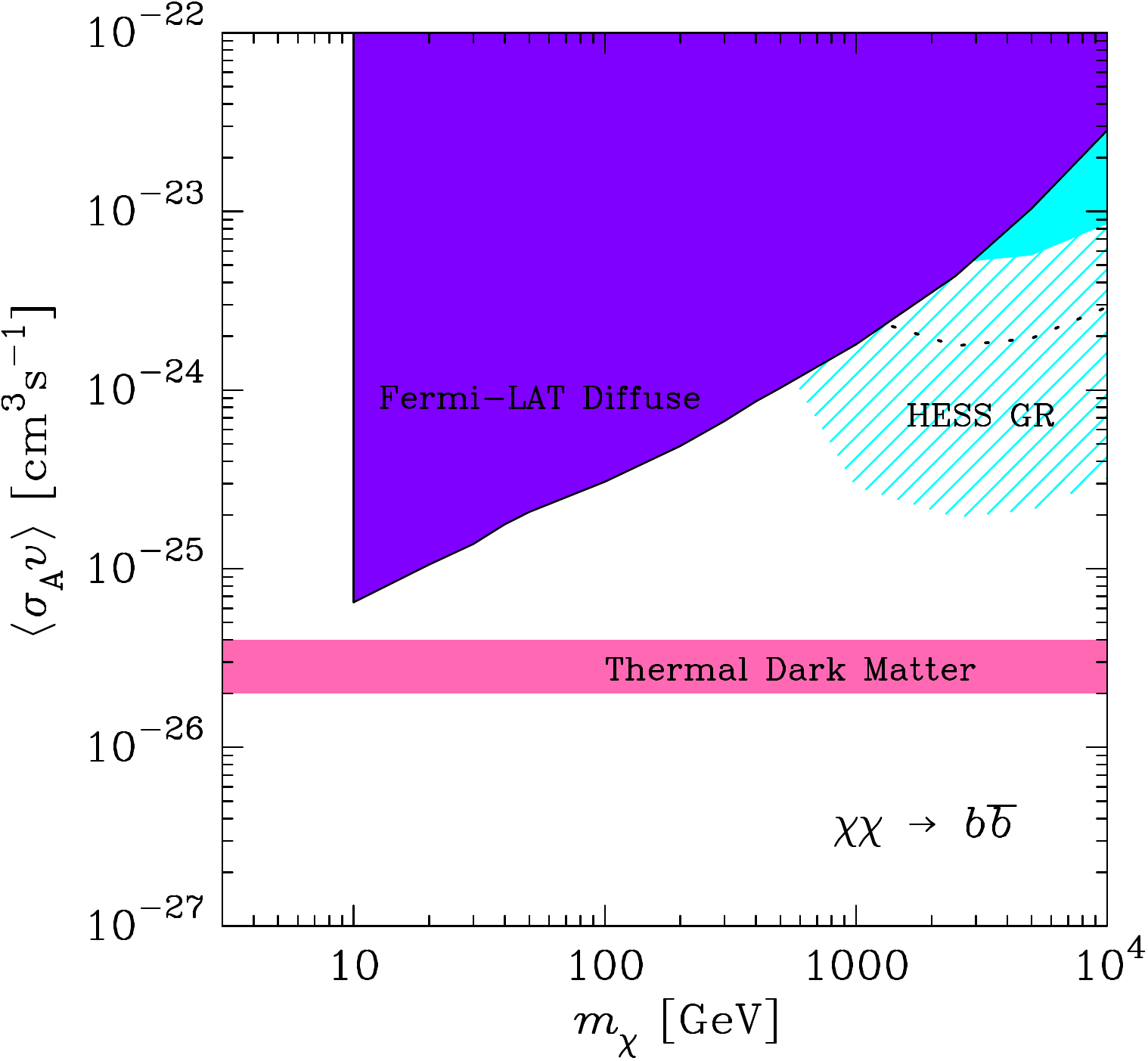}\\
 \includegraphics[width=2.2truein]{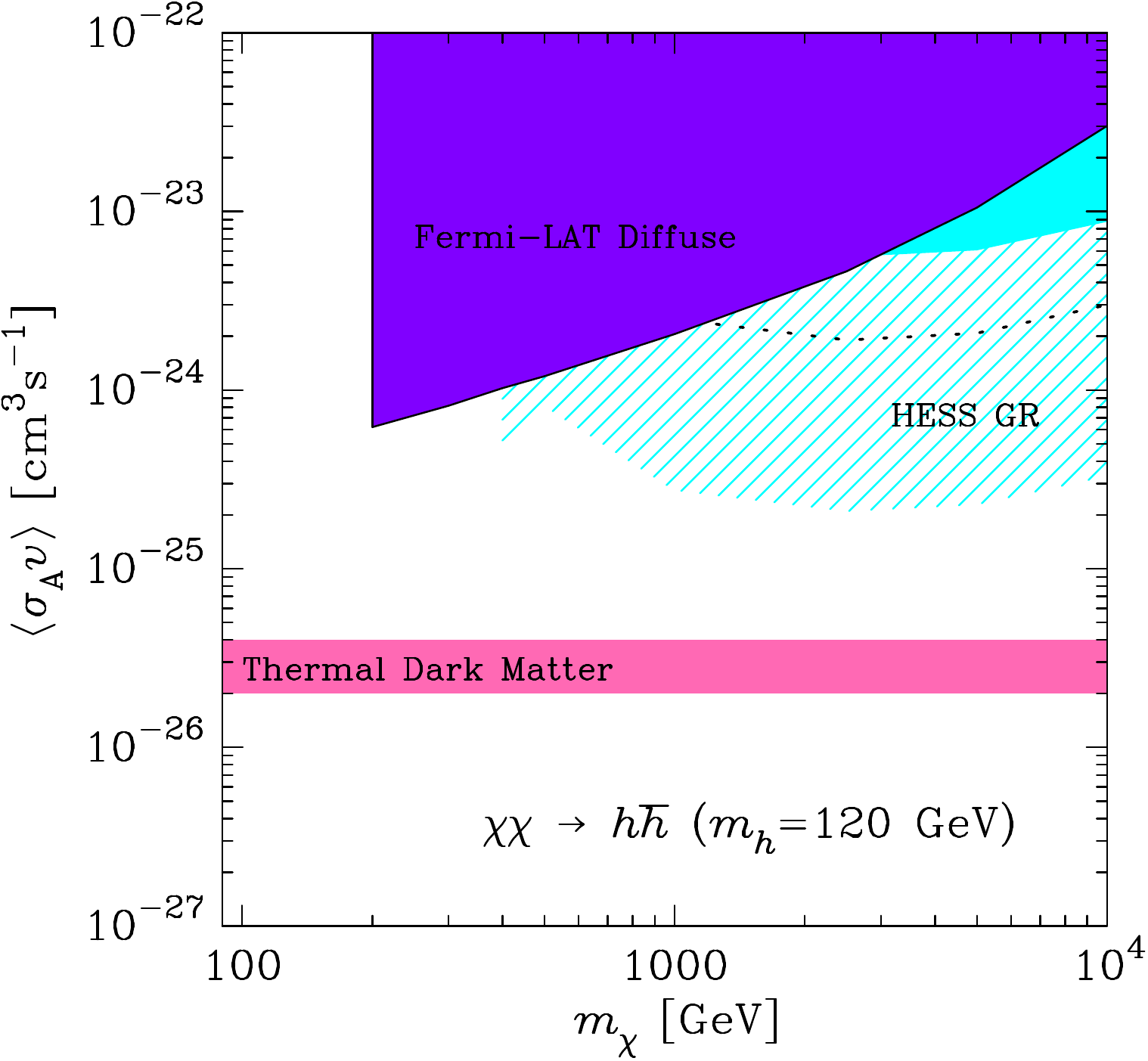}
 \includegraphics[width=2.2truein]{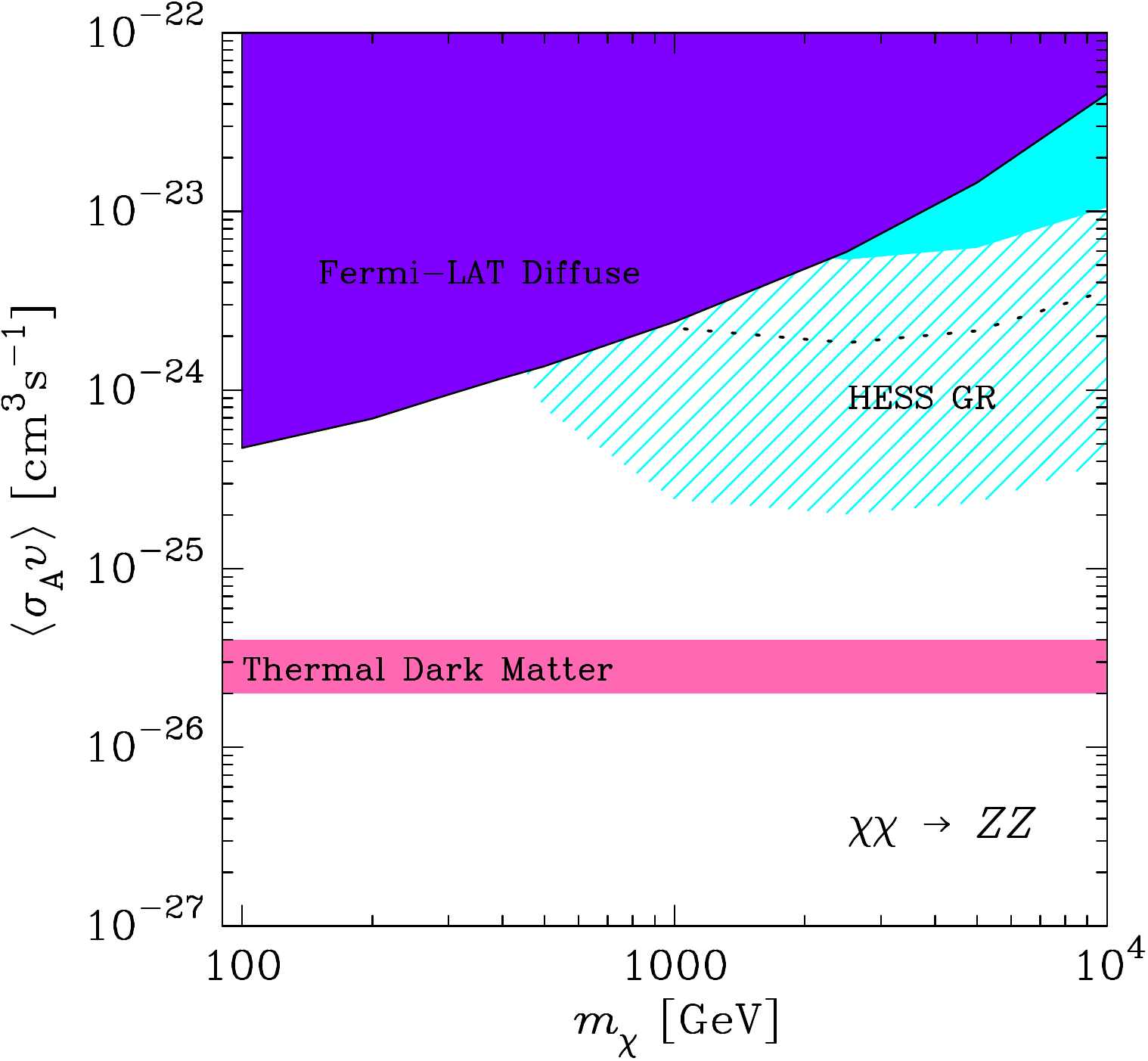}
 \includegraphics[width=2.2truein]{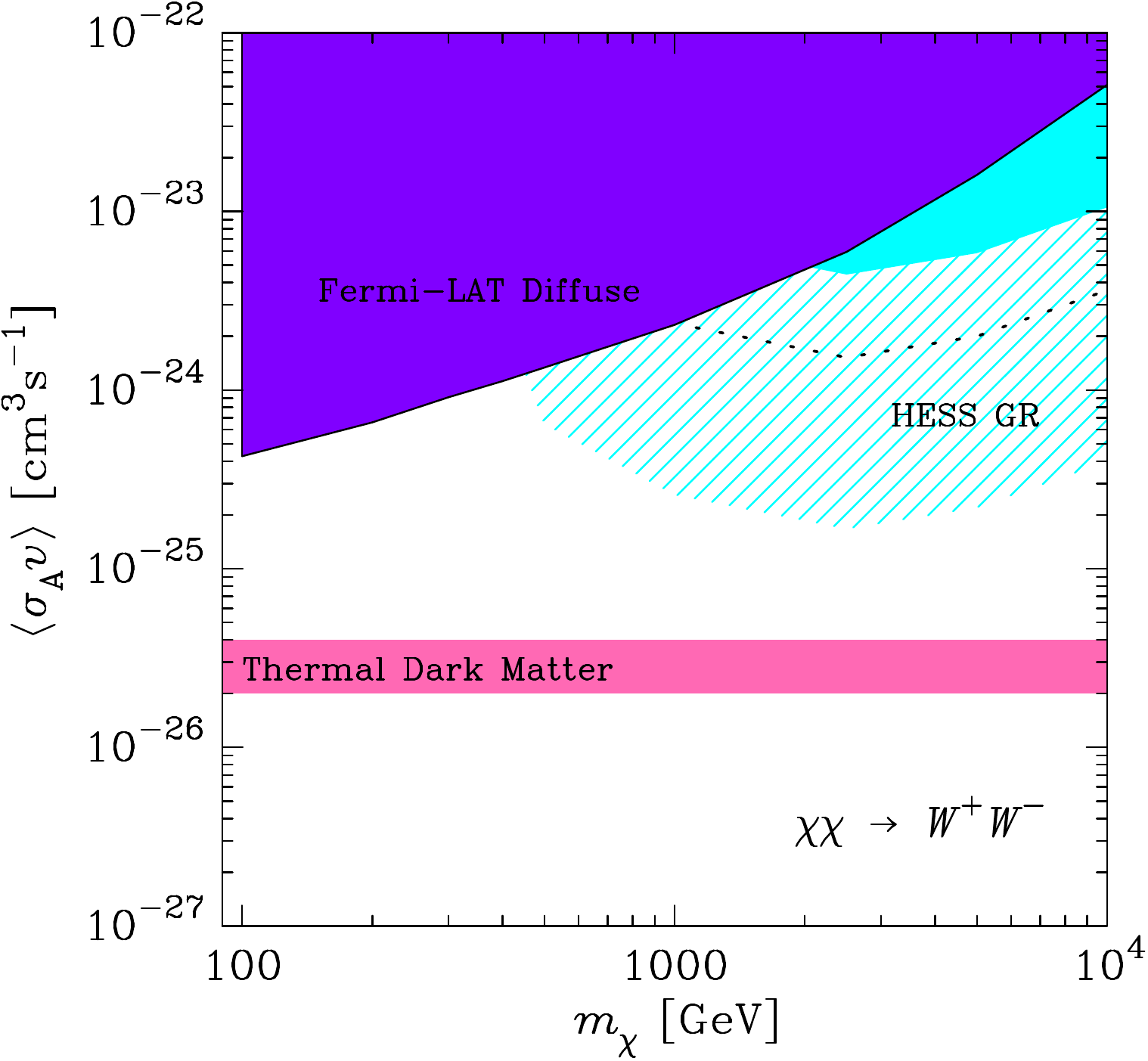}
\caption {\small Constraints on the partial cross section of
  annihilation of dark matter to quarks, leptons, gauge bosons and Higgs, for $m_h = 120\rm\
  GeV$. 
  The regions are labeled by their corresponding constraining
  observations as described in the text:  
  ``Fermi-LAT Diffuse'' from the Fermi-LAT isotropic diffuse $\gamma$-ray
  background.  The regions labeled ``HESS GR'' are for three different
  cases, solid cyan using the conservative Einasto profile as
  described in the text, hashed cyan from the most stringent case, a low
  mass high concentration NFW profile, and the canonical $\rho_\odot =
  0.3{\rm\ GeV\ cm^{-3}}, R_\odot = 8.5\rm\ kpc$ NFW profile as a dotted
  line for reference.  All constraints are at 95\% CL.
  \label{array_plot}}
%}
\end{figure*}

\section{Annihilation Signal}
A robust calculation of the expected final state radiation from dark
matter annihilation requires accurate quantification of the dark
matter source as well as the products in the final state $\gamma$-ray
radiation chain. The observable flux of photons per solid angle for a
specific annihilation channel with cross section $\langle\sigma_A
v\rangle$ is
\begin{equation}
  \frac{d\Phi_\gamma}{dE} = \frac{\langle\sigma_A v\rangle}{2}\
  \frac{\mathcal{J}_{\Delta \Omega}}{\rm J_0} \ 
  \frac{1}{\Delta \Omega_{\rm obs} m_\chi^2} \frac{dN_\gamma}{dE},
\label{fluxequation}
\end{equation}
where ${\mathcal{J}_{\Delta\Omega}}/{\rm J_0}$ is the normalized
integral of mass density squared of the dark matter in the field of
view, $dN_\gamma/dE$ is the $\gamma$-ray spectrum per annihilation,
$m_\chi$ is the dark matter particle mass, and $\Delta\Omega_{\rm
  obs}$ is the observational solid angle in steradians.  This
relationship is dependent on the astrophysically-inferred dark matter
density contribution in the field of view of an observation
$\mathcal{J}_{\Delta\Omega}$, and the particle physics of the expected
photon flux for a specific particle candidate annihilation mode,
${dN_\gamma}/{dE}$.  Note that ${d\Phi_\gamma}/{dE}$ is the total
photon number flux per unit energy per unit steradian for a full sky
observation, and when compared to the total photon count of the
Fermi-LAT observation with $|b|>10^\circ$, must be scaled to the field
of view of that observation, $\Delta\Omega_{\rm obs} = 10.4\ \rm
sr$.\footnote{An earlier version of this paper incorrectly omitted
  this geometrical factor. We thank Julie McEnery, Michael Gustafsson,
  and the Fermi Collaboration for helping resolve this issue.}  The
Fermi-LAT diffuse isotropic background spectrum observation as well as
examples of the isotropic diffuse signal for both the Galactic and
extragalactic contributions are shown in Fig.~\ref{flux_plot}.  We
describe our methods in detail below.

\subsection{Particle Annihilation Event Modeling}
\label{pythia}
For dark matter annihilations to two-body standard model final states,
in order to obtain the average number of photons above a given energy
per annihilation event and the differential number of photons at a
given energy per annihilation event $dN_\gamma/dE$, we use Pythia
6.4~\cite{Sjostrand:2006za} to simulate both photon radiation off of
charged particles as well as decays of particles such as the
$\pi^{0}$. Specifically, we run Pythia to simulate an $e^+$-$e^-$
collision at a center of mass energy of $2m_{\chi}$ through a $Z'$ to
a final state that corresponds to the annihilation products of the
dark matter. For certain final states such as $gg$ we use an s-channel
$h$ rather than a $Z'$, as the corresponding decay process is already
implemented in Pythia and does not need to be added by hand. For the
$hh$ final state we add a new decay channel for the $Z'$. We switch
off initial state radiation such that all photons are emitted either
radiatively off the final state particles or the decays of unstable
particles such as mesons. We use the default Pythia cutoff values for
photon emission: quarks are assumed not to radiate below a GeV, and
leptons below $100~{\rm keV}$. We turn on the decays of particles
which are not decayed with the default Pythia settings, such as muons,
charged pions and kaons. Using a large sample of events for each final
state and each value of $m_{\chi}$, the number of photons in the final
state above a given energy are counted and averaged over the number of
events, yielding the average number of photons above a given energy
per annihilation event.

For dark matter annihilations to final states such as $4e$ and $4\mu$
through intermediate light scalars, we utilize the formulas given in
Appendix A of Ref.~\cite{Mardon:2009rc}. For the four-$e$ final state,
all photons originate from final state radiation off of the
electrons. For the four-$\mu$ final state, photons can originate from
final state radiation off of the muons, as well as from final state
radiation off of electrons produced by the decay of the muons. Since
dark scalars are by assumption light, the energy range over which
photons can be radiated is narrow in the rest frame of the scalar.
While in the center of mass frame of the dark matter annihilation some
of these photons can still carry a significant fraction of the
available energy, the average number of hard photons per annihilation
event is significantly reduced in comparison to direct annihilation to
a two-body standard model final state.

\subsection{Cold Dark Matter Halo Models}
Thermally-produced dark matter particle candidates detectable in
$\gamma$-rays are CDM particles whose primordial velocities are only
determined by the initial gravitational perturbations.  We therefore
only consider dark matter halo models that arise in CDM cosmologies.
In particular, density profiles of approximately the canonical
Navarro-White-Frenk (NFW) profile are expected for the Milky Way (MW)
dark matter halo~\cite{Navarro:1996gj}, with profiles of the form
\begin{equation}
\rho_{\rm NFW}(r)=\rho_s \left(\frac{r}{r_s}\right)^{-1} 
               \left(1+\frac{r}{r_s}\right)^{-2},
\end{equation}
where $\rho_s$ is the characteristic density and $r_s$ is the scale
radius.  Halos in simulations exhibit scatter about this
profile~\cite{Diemand:2004wh}.  We employ the NFW profile in order to
allow for comparison to other work.  However, higher resolution
simulations of CDM halo formation have revealed a softening of the
profile power law with decreasing radius that deviates slightly but
significantly from NFW, with a logarithmic slope that decreases with
radius~\cite{Stadel:2008pn,Navarro:2008kc}. Called an Einasto profile,
it is of the form
\begin{equation}
\rho_{\rm Einasto}(r)=\rho_s
\exp\left[-\frac{2}{\alpha_E}\left(\left(\frac{r}{r_s}\right)^{\alpha_E}
    -1\right)\right],
\label{rho_einasto}
\end{equation}
where $\alpha_E$ is fit by simulations to be roughly $\alpha_E \approx 0.17\pm
0.02$.  This profile shows this self-similar behavior to the
resolution limit of the simulations at 100 kpc/$h$.

We choose conservative models the of dark matter halo profile in
calculating $\mathcal{J}_{\Delta\Omega}$ that are consistent with CDM
in addition to observational constraints of the MW halo.  The MW CDM
halo profile has been constrained with dynamical measurements by
Klypin et al.~\cite{Klypin:2001xu} and Battaglia et
al.~\cite{Battaglia:2005rj} using NFW-type halo profiles.  Recent
results by Catena \& Ullio (CU)~\cite{Catena:2009mf} have applied such
dynamical constraints to NFW as well as Einasto profiles of the MW
with a Markov-Chain Monte Carlo parameter exploration of the
degeneracies in fit parameters in these models, providing the local
(solar) density estimates of the MW halo as well as the global
profiles for these halos.  The results of CU allow using not only minimal
models in MW halo mass, but variations in other parameters as well.

There exists an overwhelming amount of results from verified numerical
simulations that halos in CDM cosmologies generally form cusped
profiles of the form exemplified by the NFW and Einasto
profiles~\cite{Navarro:1996gj,Moore:1997sg,Maccio':2006nu,Heitmann:2004gz}.
The original {\it ansatz} of CDM has been shown to be consistent with
a large range of scales of observations, including the cosmic
microwave background~\cite{Komatsu:2010fb}, large scale clustering of
galaxies and dark matter, $\sim$100 Mpc/$h$, in the Sloan Digital Sky
Survey~\cite{Eisenstein:2005su}, to the small scales of the
intergalactic medium measured in the Lyman-$\alpha$ forest, $\sim$50
kpc/$h$~\cite{Seljak:2006qw,Abazajian:2006yn,Viel:2007mv}.  We do not
consider isothermal and Burkert profiles, since cored profiles of
these forms never arise in CDM halo formation simulations of halos of
the mass of the MW.  We also do not consider the more shallow-cusped
``Kravtsov'' profiles originally from Ref.~\cite{Kravtsov:1997dp}
since following work by the same authors found sufficient force
resolution but insufficient mass resolution in the original
simulations, leading to the spurious shallower
profiles~\cite{Klypin:2000hk}.

\subsection{Galactic Center Observations}
For comparison, we include constraints from observations towards the
center of our Galaxy.  The integrated mass density squared along
line-of-sight $x$ towards Galactic coordinates $(b,\ell)$ from the GC
is
\begin{equation}
  \mathcal{J}(b,\ell) = {\rm J_0} \int_{x_{\rm min}}^{x_{\rm
      max}}{\rho^2\left(r_{\rm gal}(b,\ell,x)\right)dx}, 
  \label{jcal}
\end{equation}
where ${\rm J_0} \equiv 1/\left[8.5\ \rm kpc \left(0.3\ GeV\
    cm^{-3}\right)^2\right]$ is a normalization that makes
$\mathcal{J}$ unitless and cancels in final expressions for
observables, and 
\begin{equation}r_{\rm gal}(b,\ell,x) = \sqrt{R_\odot^2-2x R_\odot
  \cos(\ell)\cos(b)+x^2}.
\end{equation}
The lower and upper limit of the
integration $x_{\rm min}$, $x_{\rm max}$ are set by whether it's the
MW or an extragalactic halo.  The limits chosen are somewhat arbitrary
since galactic halos are embedded and continuous with a network of
lower-density filamentary structures.  However, the contribution
beyond the scale radius is minimal, and we simply choose a cutoff at
the nominal virial radius for the MW.  The average of the
integrated mass density is
\begin{equation}
  \mathcal{J}_{\Delta\Omega} = \frac{1}{\Delta\Omega}\int_{\Delta\Omega} 
  {\mathcal{J}(b,\ell) d\Omega}, 
  \label{skyaverage}
\end{equation}
over the relevant sky region, where $\Delta\Omega$ is the angular size
of the observation region.  Although the NFW profile is divergent at
the GC, the average over a finite observation region is insensitive to
the exact nature of the inner cusp for the $\gamma$-ray observations
here.  We regularize the NFW profile by fixing a constant density of
the profile within $r_{\rm halo} < 10^{-10}\ \rm kpc$ to the density at
$\rho(r_{\rm halo}=10^{-10}\ \rm kpc)$ in Eq.~\ref{skyaverage}, which
has negligible effect on the integrated average.

One of the most stringent limits on dark matter annihilation arise
from observations of the MW Galactic center ridge (GR) by the High
Energy Stereoscopic System (HESS) telescope~\cite{Aharonian:2006au},
an array of atmospheric \v{C}erenkov telescopes in operation in
Namibia\footnote{http://www.mpi-hd.mpg.de/hfm/HESS/HESS.html}~\cite{Aharonian:2007km,
  Regis:2008ij, Mack:2008wu, Bell:2008vx, Bertone:2008xr,
  Meade:2009iu, Bi:2009de}.  The HESS GR observations are for an
angular region $-0.8^\circ <\ell < 0.8^\circ$ and $-0.3^\circ<b
<0.3^\circ$.  We integrate Eq.~(\ref{skyaverage}) directly, through a
Monte-Carlo integration method over the nontrivial geometry.  Note
that in HESS GR observation subtracted a background from a region
$-0.8^\circ <\ell < 0.8^\circ$ and $0.8^\circ<b <1.5^\circ$ to remove
cosmic ray backgrounds.  This also removes any concurrent dark matter
signal in the background region, which we incorporate appropriately by
integrating the signal from the background subtracted region
$-0.8^\circ <\ell < 0.8^\circ$ and $0.8^\circ<b <1.5^\circ$
numerically, also through a Monte Carlo integration method for the
arbitrary geometry.\footnote{Note that we do not adopt an
  often-employed circular area approximation for the HESS GR
  observational geometry, or a point approximation for the background
  subtraction, e.g. in \cite{Mack:2008wu, Bell:2008vx}.  The circular
  approximation underestimates the signal by approximately a factor of
  two, primarily because the observational area is overestimated while
  the signal is centrally concentrated.  The circular area
  approximation with angle $\psi = 0.8^\circ$ has $\Delta\Omega_{\rm
    circ} = 2\pi(1-\cos\psi) = 6.1\times 10^{-4}\rm\ sr$ while the
  HESS GR observation was over an area $\Delta\Omega_{\rm HESS} =
  2.9\times 10^{-4}\ \rm sr$.  The background approximation of a point
  subtraction at $\psi=0.8^\circ$ overestimates the background
  integral, yet this is partially offset by the smaller area of the
  background $\Delta\Omega_{\rm HESS,bg} = 3.4\times 10^{-4}\ \rm sr$
  than in the point approximation.}  The full Monte Carlo integration
particularly affects the Einasto profile case due to the
shallower Einasto profile which falls off more slowly than NFW in the
background region, and therefore increases the background subtraction.

We calculate extremal cases for $\mathcal{J}_{\Delta\Omega,\rm HGR}$
towards the HESS GR field of view, such that
\begin{eqnarray}
\mathcal{J}_{\Delta\Omega,\rm HGR} = 
\begin{cases}
555\quad&\text{CU Einasto Minimum}\cr
1050\quad&\text{Battaglia NFW Low Mass}\cr
2500\quad&\text{CU NFW Low $c$}\cr
13100\quad&\text{Battaglia NFW High Mass}\cr
14700\quad&\text{CU NFW High $c$}.
\end{cases}
\label{j_hessgr}
\end{eqnarray}
For the minimal Einasto case, we use a softer conservative value of
the log-slope that is consistent with the fits of CU, with $\alpha_E =
0.22$, $r_s = 21\rm\ kpc$, $r_\odot = 8.28\rm\ kpc$ and $\rho_\odot =
0.385\rm\ GeV\ cm^{-3}$ which are the values we adopt as our
conservative minimal case of the MW halo profile in the HESS GR field
of view.  Our adopted maximally stringent case is the NFW profile from
CU which has the highest concentration, $c$.  High concentration
enhances the signal toward the GC, yet is anti-correlated with halo
mass~\cite{Catena:2009mf}, therefore corresponding to a low mass MW
halo, with $c=24.6$ and $M_{\rm vir}=1.23\times 10^{12}\ M_\odot$.
The CU NFW low $c$ case has $c=13.9$ and $M_{\rm vir}=1.86\times
10^{12}\ M_\odot$ (we do not employ this model in our constraints).
Both CU NFW profiles adopt $r_\odot = 8.28\rm\ kpc$. The high and low
mass NFW cases from Battaglia et al.~\cite{Battaglia:2005rj}
correspond to the 68\% CL limits of the MW halo model: $M_{\rm vir} =
0.6 \times 10^{12}\ M_\odot$ (low mass), $M_{\rm vir} = 2.0 \times
10^{12}\ M_\odot$ (high mass).  Both of these cases have other
parameters fixed at $R_{\rm vir} = 255\rm kpc$, $c=R_{\rm
  vir}/r_s=18$, and $r_\odot = 8\rm\ kpc$. The Battaglia et al. NFW
extremal profiles and corresponding GC $\mathcal{J}$ are consistent
with those from CU, see Eq.\ (\ref{j_hessgr}).  The CU and Battaglia
et al. NFW profiles are also consistent with and comparable to the
high and low mass MW halo models of Klypin et
al.~\cite{Klypin:2001xu}.  In order to allow for comparisons with
previous work, we also evaluate the constraints for the HESS GR for
the case of the canonical NFW halo with $\rho_\odot = 0.3\ \rm GeV\
cm^{-3}$, $R_\odot = 8.5\rm \ kpc$, $r_s = 20\rm\ kpc$, where
$\mathcal{J}_{\Delta\Omega,\rm HGR} = 1620$, and the constraint in
parameter space is plotted as a dotted line in
Fig.~\ref{array_plot}.

Our results are less stringent for the Einasto case compared to
Refs.~\cite{Bertone:2008xr,Meade:2009iu} because the signal used in
that work did not remove the signal from the background subtraction
region.  Note also that our minimal case of a ``CU Einasto Minimum''
is significantly more stringent in its constraints than the adopted
values in, {\it e.g.}  Refs.~\cite{Mack:2008wu,Bell:2008vx} which
adopted inaccurate MW profiles from Ref.~\cite{Kravtsov:1997dp} which
had known insufficiencies in numerical mass resolution as discussed
above.  We do not include here constraints from the smaller region
HESS GC observations presented in Ref.~\cite{VivierHESSGC} since they
are weaker than the HESS GR observations for nearly all channels and
regions in parameter space~\cite{Bertone:2008xr}.

\subsection{The Dark Matter Signal in the Isotropic Diffuse Background}
\label{dmhalo}
The MW halo itself contributes to the isotropic background in an
irreducible fashion due to the smooth and unresolved portion of the MW
halo.  We use a very conservative limit in the required contribution
of the MW Galactic halo to the isotropic diffuse background.
Specifically, we assume that the entire diffuse MW signal is at
minimum the value opposite the GC from a smooth halo, with a minimal
substructure boost. We take that $\mathcal{J}_{\Delta\Omega}$ is given
by the anti-GC $\mathcal{J}(0,180^\circ)$ averaged over the full
diffuse observation region $\Delta\Omega$, $\mathcal{J}_{\rm
  Iso}\simeq 0.62$.  The signal towards the most Galactic-centric
portion of the Fermi-LAT diffuse observation region is over two orders
of magnitude larger than that away from the GC,
$\mathcal{J}(10^\circ,0)\simeq 78$.  Therefore, our approach is
extremely conservative, and leads to a robust bound since the removal
of a spatial dependence in the Fermi-LAT diffuse spectrum may remove
spatially-dependent components of the predicted signal.

A cold dark matter halo necessarily has enhanced annihilation from
substructure in the dark matter halo.  Early work had found MW
substructure to contribute to boost factors of the total MW
annihilation signal at over two orders of magnitude greater than the
smooth halo contribution~\cite{CalcaneoRoldan:2000yt}.  More recent
work from the Aquarius simulations find similar boost factors at 232
times the smooth contribution~\cite{Springel:2008zz}.  The minimal
level of enhancement from numerically resolved substructure has been
found to be a factor of $\sim$2 from the Via Lactea and Via Lactea II
simulations~\cite{Diemand:2006ik,Kuhlen:2008aw}.  However,
substructure is expected to exist in CDM halos at potentially over
twenty orders of magnitude in mass scale below a MW parent halo.
Therefore, we go beyond the approximations from resolution
limitations.

\begin{figure}
%\FIGURE{
%\includegraphics[width=6.truein]{bbbar_0datfermi_comp_labels2.pdf}
\includegraphics[width=3.2truein]{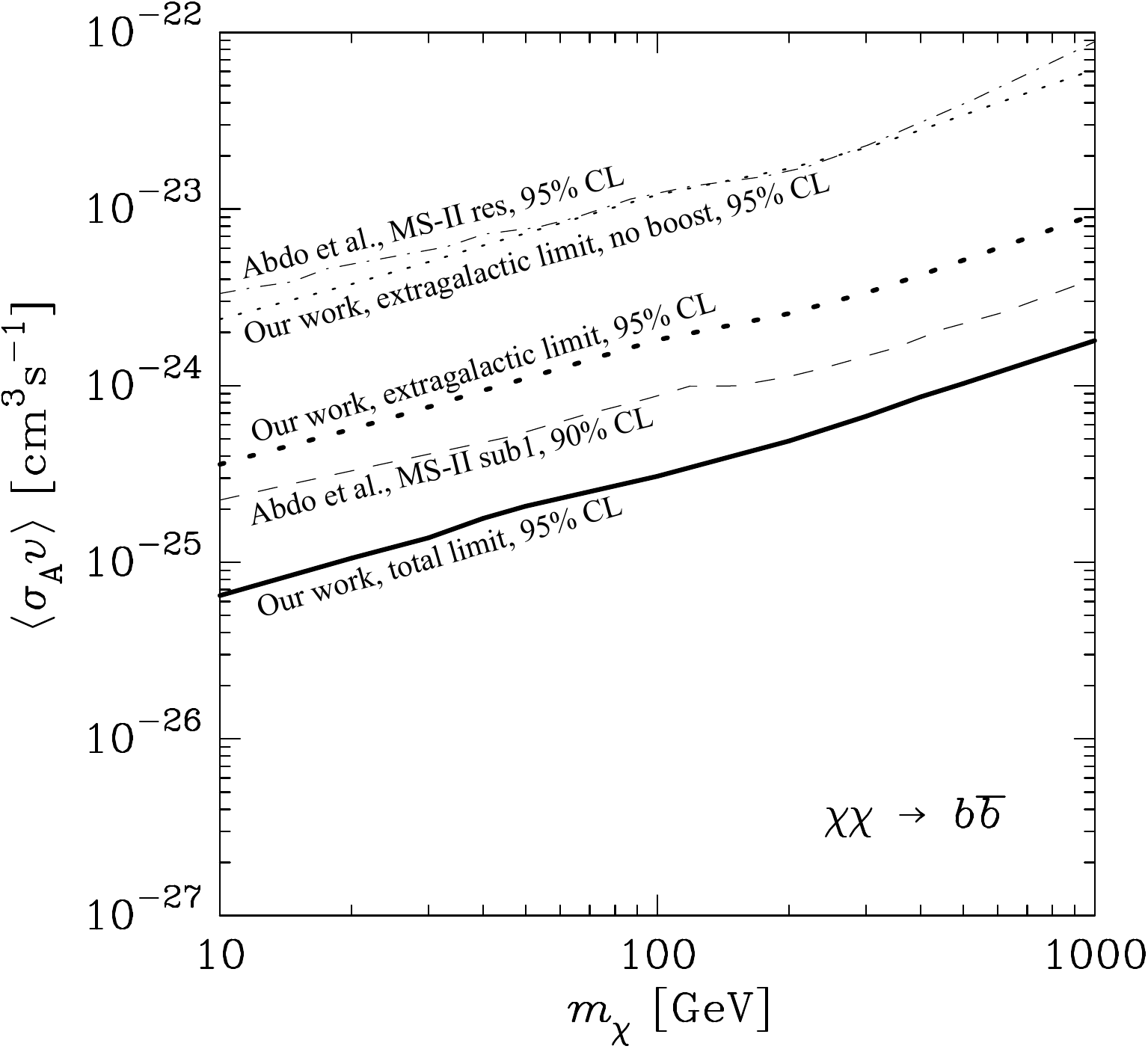}
\caption {\small Shown is a comparison of our work to the Fermi-LAT
  collaboration, Abdo et al.~\cite{Abdo:2010dk} constraints from the
  extragalactic signal in the diffuse extragalactic background from
  dark matter annihilation into the $b\bar b$ final state.  Our limits
  from the extragalactic background with not boost enhancement of
  annihilation are comparable to and consistent with that of the Abdo
  et al.  ``MS-II Res'' model due to similar chosen halo mass cut-off
  values.  Our total 95\% CL limit shown here and used in
  Fig.~\ref{array_plot} comes from Galactic and extragalactic
  contributions, with a conservative boost factor, as described in the
  text.  Shown also is our 95\% extragalactic limit alone, and the
  conservative model in Abdo et al. of ``MS-II
  sub1.'' \label{comparison_plot}}
%}
\end{figure}

In order to provide a conservative estimate of the enhancement of
annihilation due to unresolved substructure in the Fermi-LAT
observation, we employ the analytic density probability distribution
function (PDF) proposed by Kamionkowski \&
Koushiappas~\cite{Kamionkowski:2008vw}. This method was recently
applied to the enhancement of annihilation signals in MW-type halos
with a calibration of the PDF with the Via Lactea II (VL-II)
simulations~\cite{Kamionkowski:2010mi}.  That work found typical boost
factors for the full MW halo are approximately $17$.

In order to
quantify the lower limit of the boost factor, we extremize the PDF to
the lowest values consistent with the CDM halo calibration.  The mean
halo density from the PDF is given by
\begin{eqnarray}
\bar\rho(r) &=& \int_0^{\rho_{\rm max}}{\rho P(\rho) d\rho} \\
&=& f_s\rho_h +\nonumber \\
&\ & (1-f_s)\rho_h
\begin{cases}
\frac{1+\alpha}{\alpha}\left[1-\left(\frac{\rho_{\rm max}}{\rho_h}\right)^{-
\alpha}\right];\ \ \alpha = 0,\\
(1-f_s)\rho_h \ln \left(\frac{\rho_{\rm max}}{\rho_h}\right); \ \ \alpha\neq 0.
\end{cases}
\label{rhobar}
\end{eqnarray}
The fraction $f_s$ of the halo volume is filled with a smooth dark
matter component with density $\rho_h$.  The power law tail is
calibrated to simulations to be
\begin{equation}
(1-f_s) = 7\times 10^{-3}\left(\frac{\bar\rho(r)}{\bar\rho(r=\rm 100\ kpc)}\right)^{-0.26},
\end{equation}
for radii greater than 20 kpc, where the overwhelmingly dominant
portion of the substructure boost arises.  The normalization of this
tail is lower at smaller radii, $\lesssim 20\rm\ kpc$, but the
contribution to the cumulative boost from these radii is
minimal. Simulations find that $\alpha = 0\pm 0.1$. 

%\FIGURE{ 
\begin{figure*}
\includegraphics[width=2.2truein]{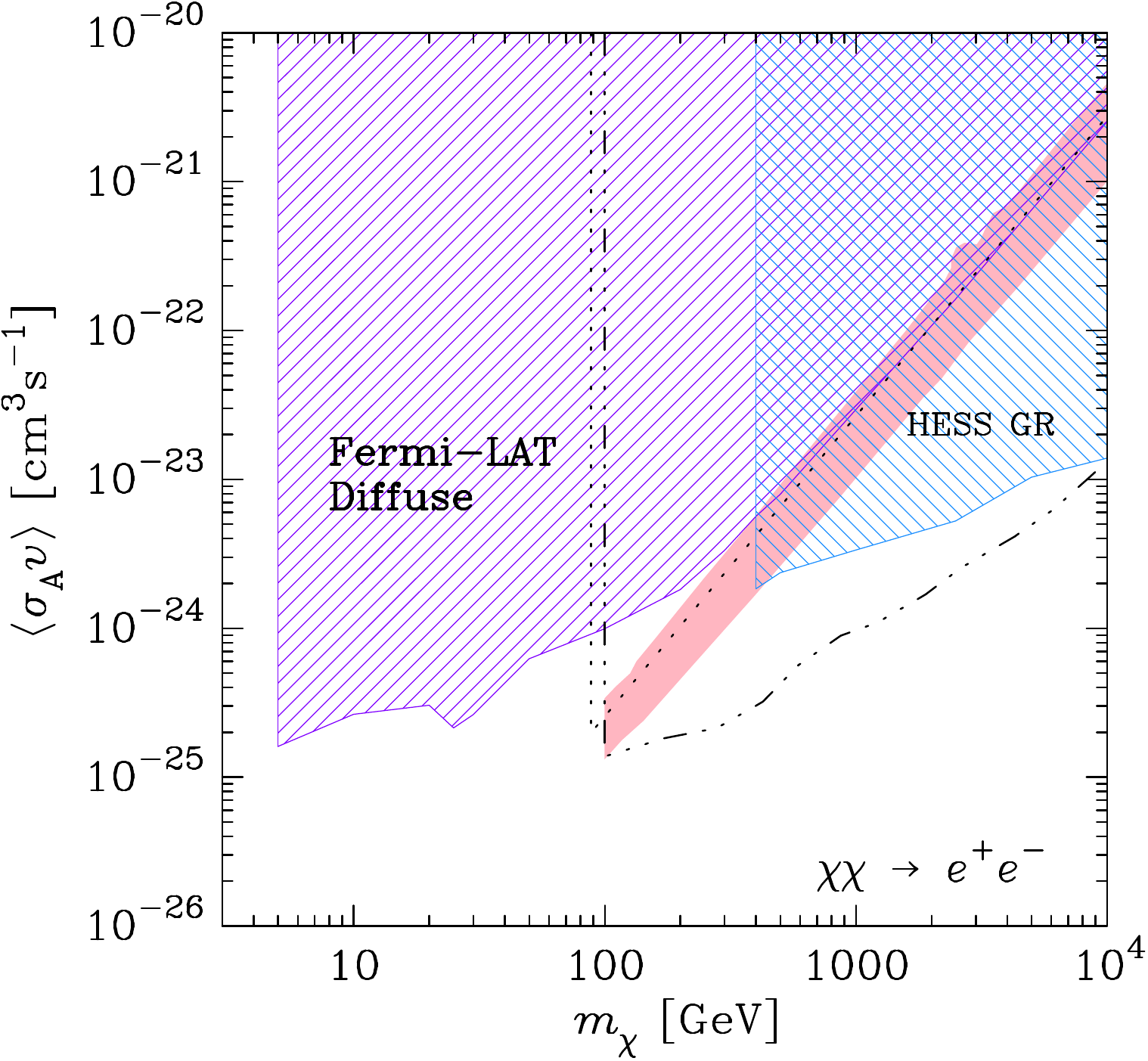}
\includegraphics[width=2.2truein]{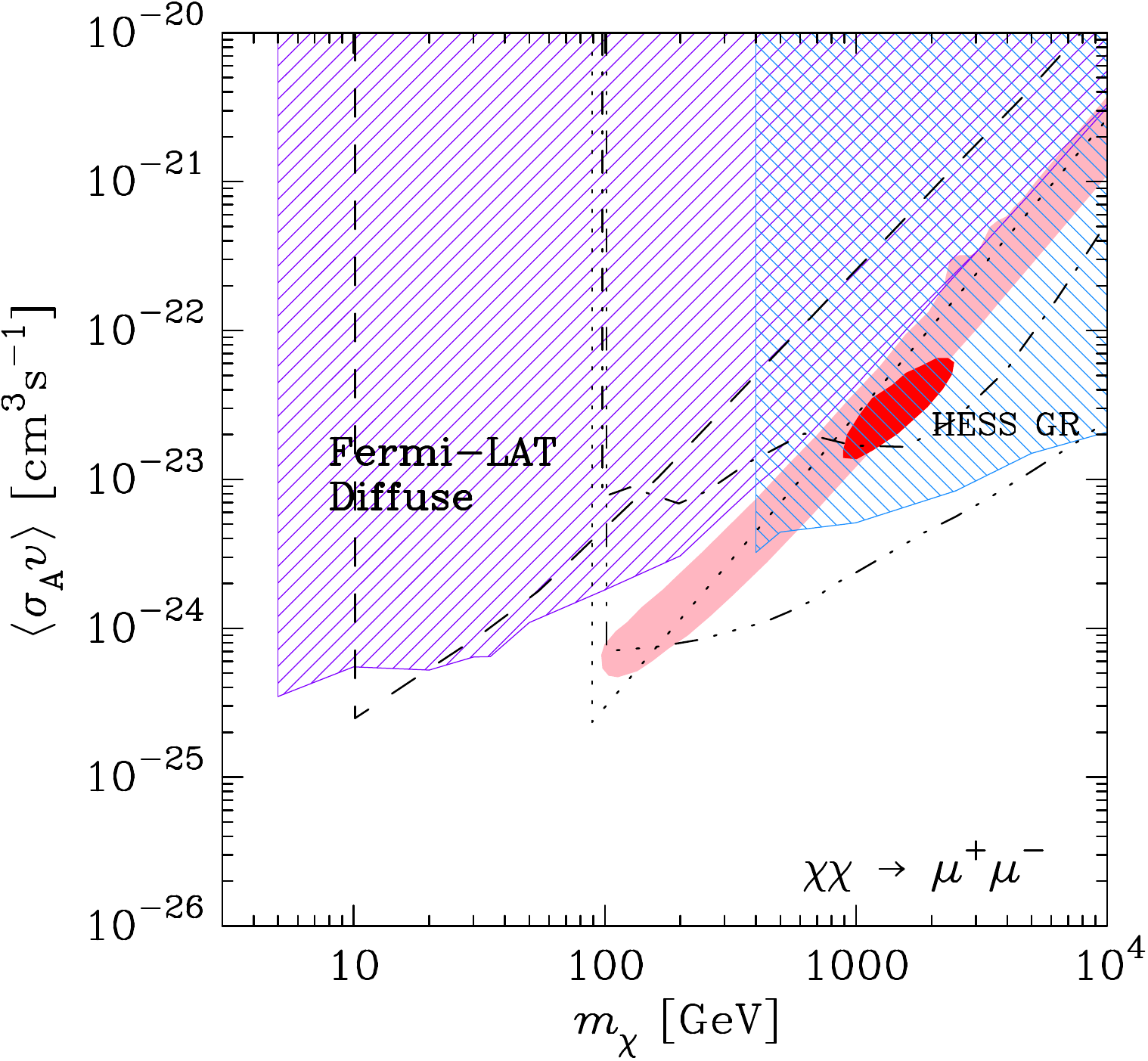}
\includegraphics[width=2.2truein]{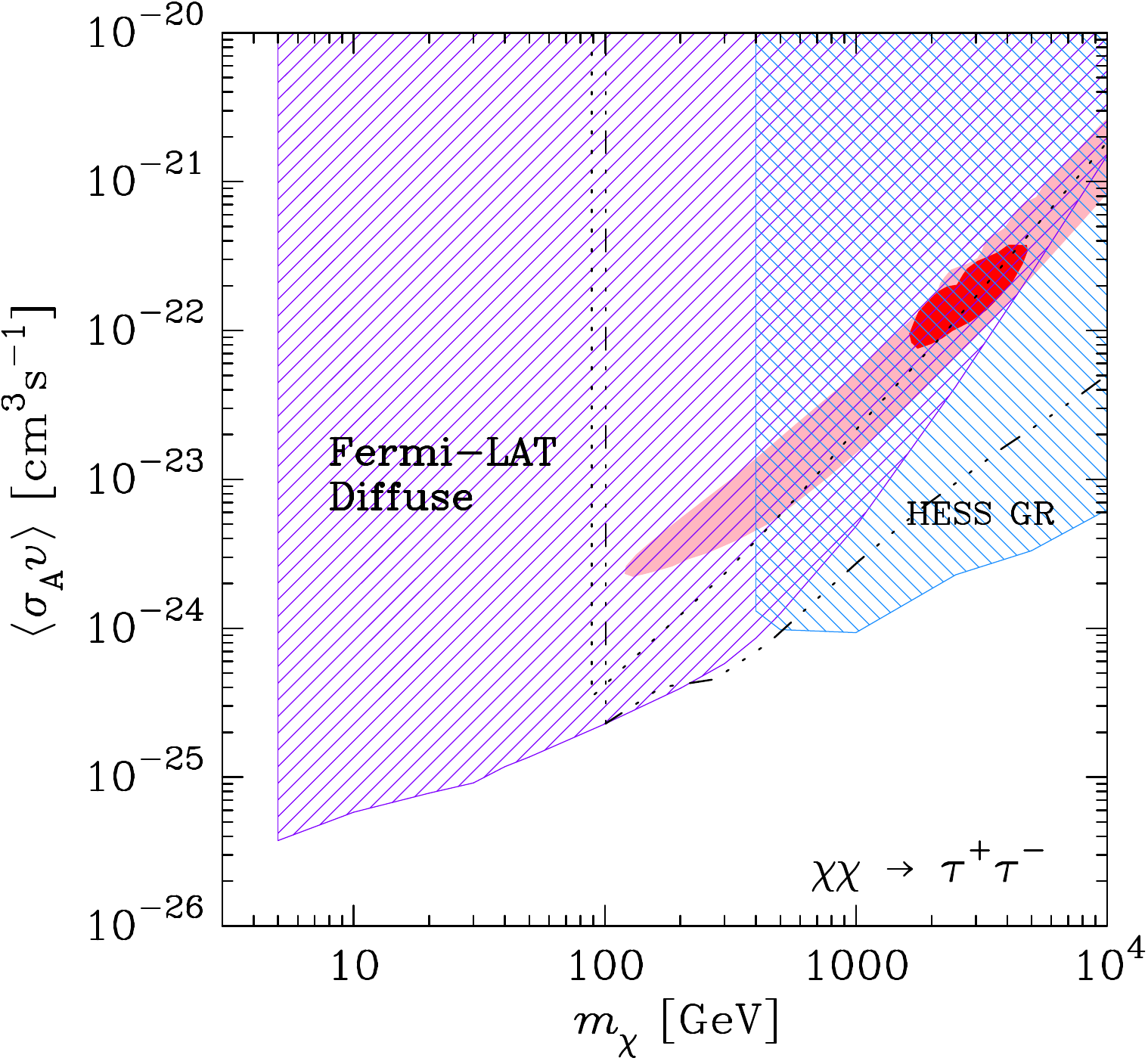}
\caption {\small Shown are constraints on regions of parameter space
  in annihilating dark matter models consistent with interpretations
  of the PAMELA positron excess (light pink region in all panels), and
  feature in the Fermi-LAT $e^+/e^-$ spectrum (red region in
  $\mu^+\mu^-$ and $\tau^+\tau^-$ panels), from
  Ref.~\cite{Meade:2009iu} (all 99\% CL).  All MW
  constraints and signals are for Einasto profiles, and at 95\% CL.  
  Our analysis of    the Fermi-LAT isotropic diffuse background is
  shown along with  
  our reanalysis of constraints from HESS observations of the Galactic
  Ridge (labeled HESS GR).  We show several other recent constraints
  for comparison: ICS radiation constrained by Fermi-LAT
  data from the $3^\circ\times 3^\circ$ Galactic center exclude the region within the
  triple-dot-dashed regions in all panels~\cite{Cirelli:2009dv}; 
  Galactic radio synchrotron
  observations exclude the region within the dotted line, in all
  panels~\cite{Meade:2009iu}; in the central $\mu^+\mu^-$ panel, we
  show the exclusion regions final state radiation constraints in 
    Fermi-LAT observations of Draco (dashed line), and ICS
   radiation constraints from Fermi-LAT
  observations of Ursa Minor (dot-dashed
  line)~\cite{Abdo:2010ex}.  All PAMELA models with
  $m_\chi \lesssim 1\rm\ TeV$ are firmly excluded by the lack of a
  $\gtrsim 20\%$ drop in the Fermi $e^+/e^-$ spectrum below 1
  TeV~\cite{Meade:2009iu}.  
  \label{pamela_plot1}}
%}
\end{figure*}

%\FIGURE{ 
\begin{figure}
 \includegraphics[width=2.2truein]{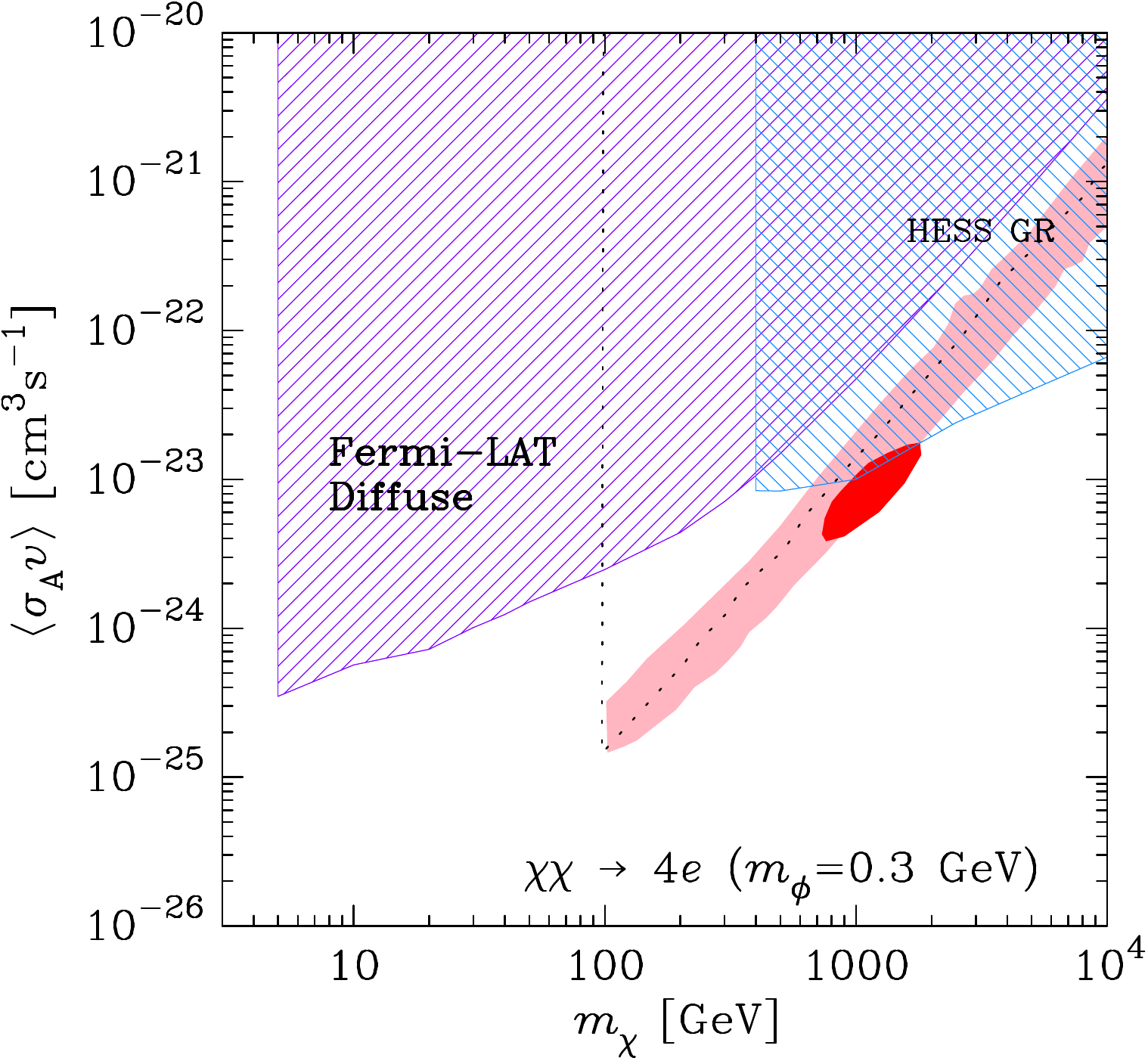}\\
 \includegraphics[width=2.2truein]{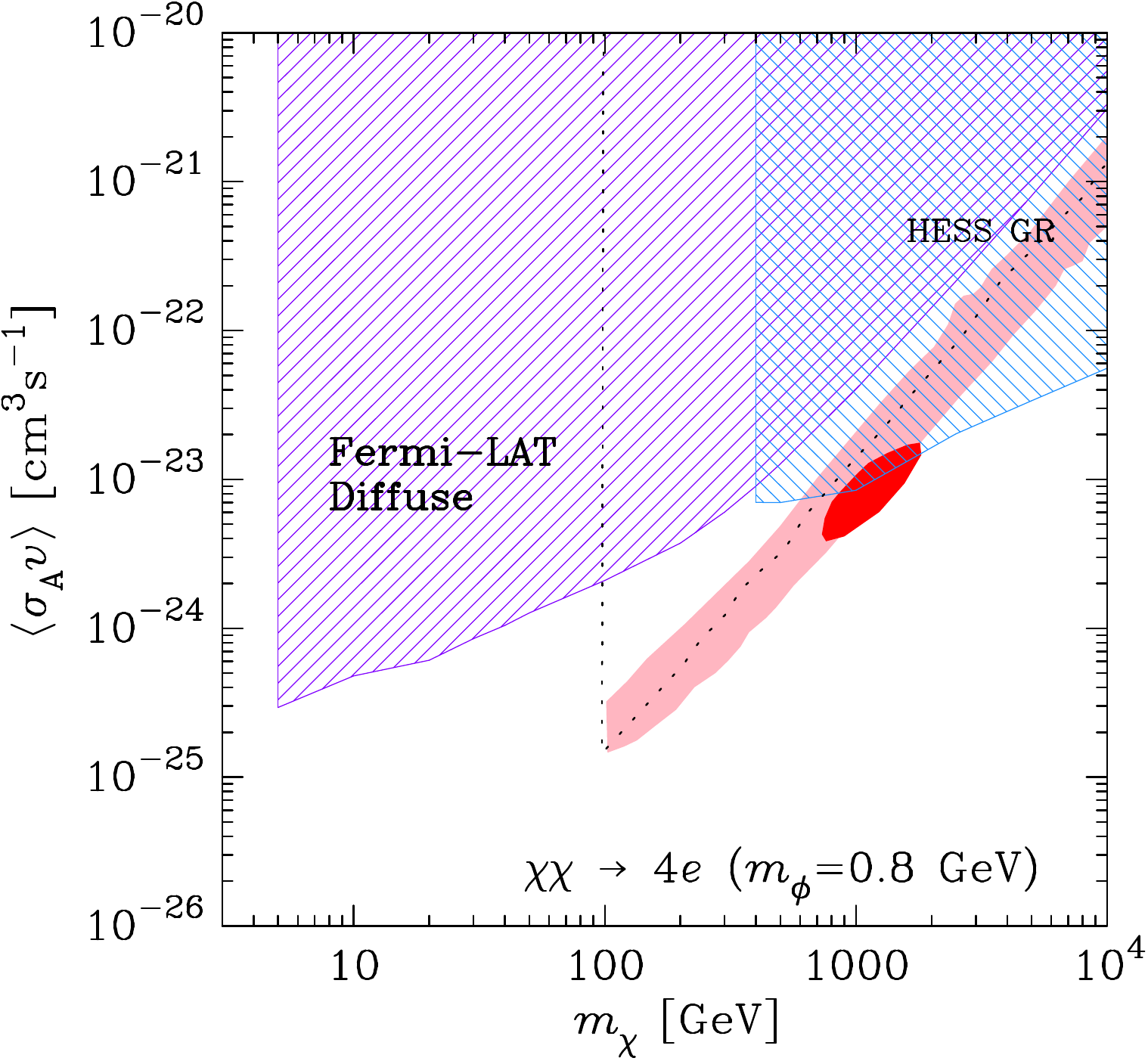}\\
 \includegraphics[width=2.2truein]{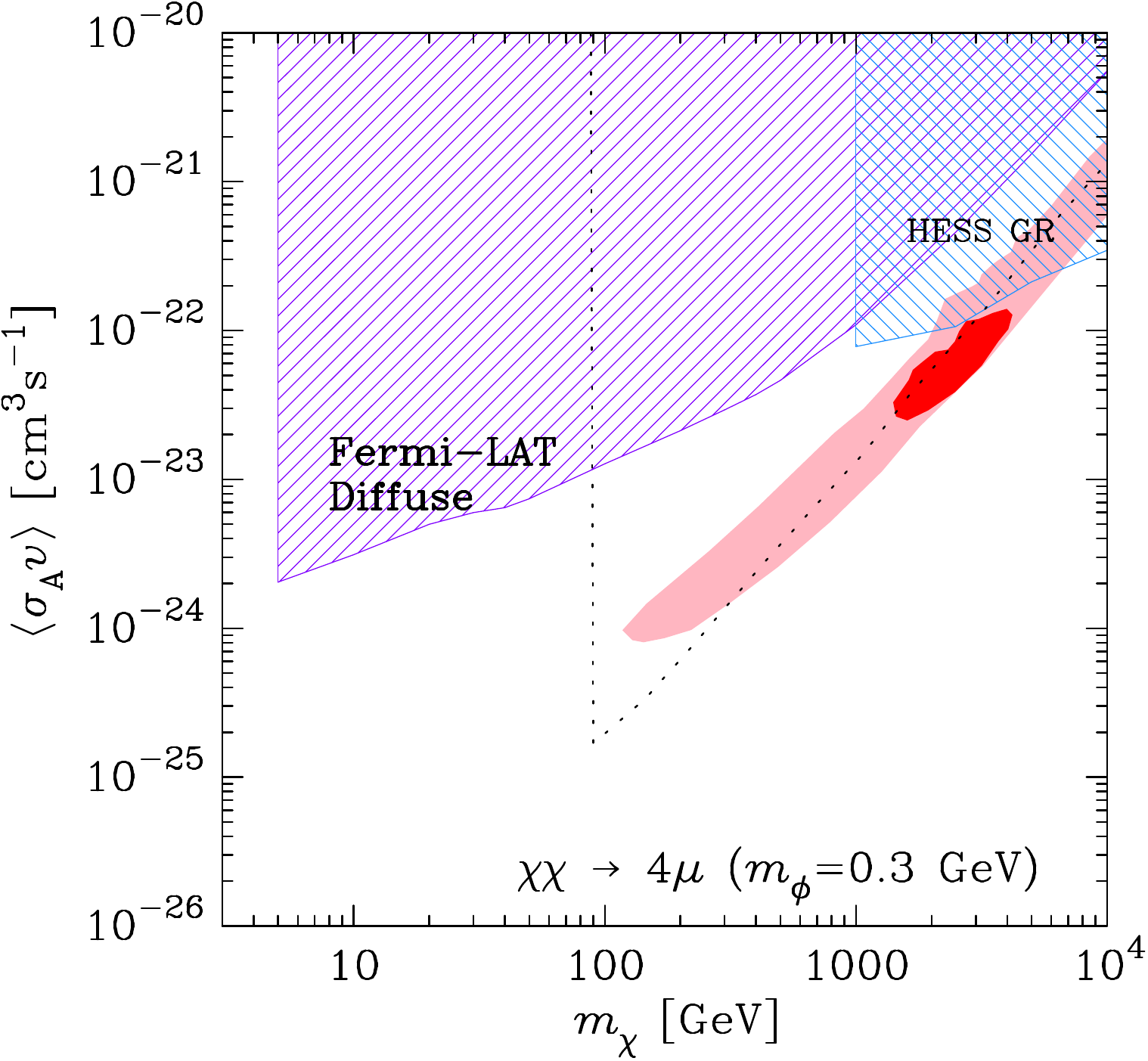}\\
 \includegraphics[width=2.2truein]{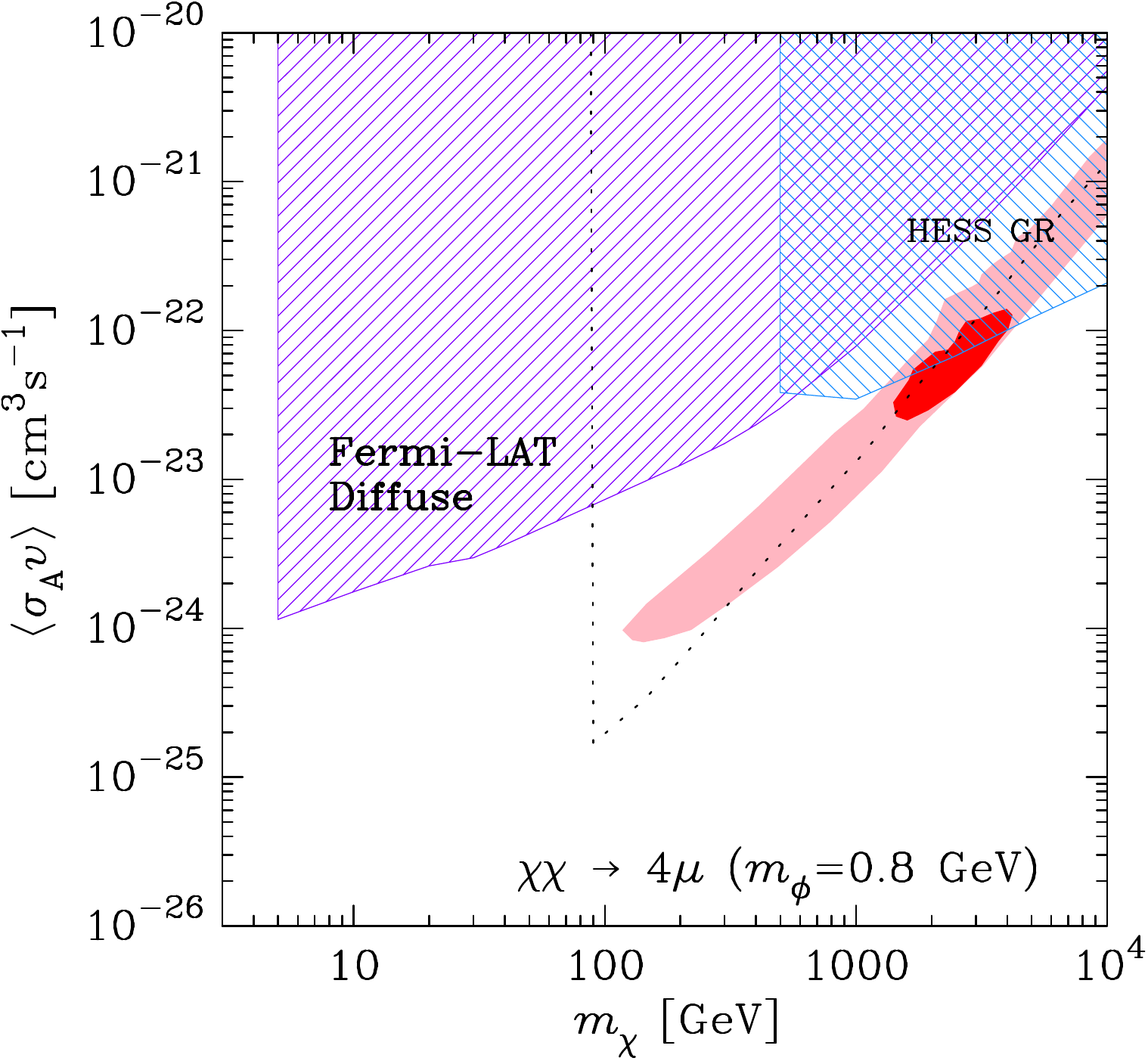}
\caption{Interpretations of PAMELA and Fermi $e^+/e^-$ with an
  intermediate dark force carrying particle $\phi$ allowing for dark
  matter annihilation into four lepton final states.  The upper
  (lower) two panels are $4e$ ($4\mu$) final states with $m_\phi =
  0.3\rm\ GeV$ and $m_\phi = 0.8\rm GeV$.  Galactic radio synchrotron
  observations exclude the region within the dotted line, in all
  panels~\cite{Meade:2009iu}.  All PAMELA models with $m_\chi \lesssim 1\rm\
  TeV$ are firmly excluded by the lack of a $\gtrsim 20\%$ drop in the
  Fermi $e^+/e^-$ spectrum below 1
  TeV~\cite{Meade:2009iu}.  \label{pamela_plot2}}
%}
\end{figure}

The definition of the annihilation boost factor including the CDM
density PDF, as a function of galactic radius is
\begin{equation}
B(r) = \frac{\int\rho^2\ dV}{\int\left[\bar\rho(r)\right]^2\ dV}  =
\int_0^{\rho_{\rm max}} P(\rho,r)
\frac{\rho^2}{\left[\bar\rho(r)\right]^2} d\rho.
\end{equation}
Explicitly, for this PDF, the boost factor is
\begin{eqnarray}
B(r) &=& f_s e^{\Delta^2}\nonumber \\
&\ & +(1-f_s) \frac{1+\alpha}{1-\alpha}\left[\left(\frac{\rho_{\rm
        max}}{\rho_h}\right)^{1-\alpha} -1\right].
\end{eqnarray}
The first term $f_s e^{\Delta^2}$ is due to the variation in the
smooth component, and since $\Delta\lesssim 0.2$ from simulations, it
contributes to the overall boost factor by only a few percent and we
ignore it.  The second term is the boost factor due to substructure.
The total boost in the galactic halo is 
\begin{equation}
B(<R) = \frac{\int_0^R B(r) \rho(r)^2 r^2 dr}{\int_0^R \rho(r)^2 r^2
  dr} \label{bofr}
\end{equation}
We take the extremal case to be where the slope of the power-law tail
is steeper than the mean found in simulations, $\alpha = 0.1$.  We
take $\rho_{\rm max}$ to be $\rho_s$ of the earliest forming halos,
which have low concentrations $c\approx 2-5$, which for centrality of
this concentration distribution we take $c=3.5$, forming at $z_c =
40$.  Note that using $\rho_s$ is extremely conservative since the
smallest subhalos likely have cuspy profiles themselves that reach
densities much higher than their scale density $\rho_s$.  The
integral, Eq.~(\ref{bofr}) must be evaluated numerically, and for our
adopted conservative case is
\begin{equation}
  B(< R_{\rm vir}) = 6.6,
\label{totalboost}
\end{equation}
for $R_{\rm vir} = 255\rm\ kpc$.  Note that this total boost is
significantly lower than typically found in analytic extrapolations of
MW halos which find boost factors of over two orders of magnitude.  We
employ this minimal boost factor, Eq.~(\ref{totalboost}), in our
signal calculations.  It should be noted that the boost in the local
(solar) region is minimal, on average, but including the boost due to
substructure can affect and improve the modeling of leptonic
cosmic-ray annihilation signals
\cite{Kamionkowski:2010mi,Cline:2010ag}.

A similar modeling to the one we adopt here of the density PDF
calibrated to cosmological volume $(100\ {\rm Mpc}/h)^3$ simulations
found larger scatter of the power law tail at mass scales approaching
MW scale halos with boost factors between 2 and
2000~\cite{Zavala:2009zr}.  The mass resolution of the Millenium
Simulation II (MS-II) employed in Ref.~\cite{Zavala:2009zr}
(particle mass $6.89\times 10^6\ M_{\odot}/h$) is approximately 2400
times less than that in VL-II (particle mass $4100\ M_\odot$), and the
MS-II results did not extend to PDF resolution of MW size halos.  This
may contribute to the increased PDF scatter for lower mass halos, and
therefore we adopt the VL-II calibrated PDF of
Ref.~\cite{Kamionkowski:2010mi}.  Note that a complete systematic
analysis of the lower limit of the substructure PDF contribution to
the boost has not been performed since some of the earliest work on
substructure boost factors, which found total boost scatter at the
level of 40\%~\cite{CalcaneoRoldan:2000yt}. Such a lower scatter is
consistent with what we find, Eq.~(\ref{totalboost}).

The cosmological contribution to the isotropic diffuse $\gamma$-ray
background from annihilating dark matter has been studied for some
time~\cite{Bergstrom:2001jj,Ullio:2002pj}.  The diffuse flux from
annihilation in cosmological extragalactic halos is
\begin{eqnarray}
  \frac{d\Phi_\gamma}{dE} &=&\frac{\langle\sigma_A v\rangle}{2}\ 
\frac{c}{\Delta\Omega_{\rm obs} H_0}\ \frac{(f_{\rm DM} \Omega_m)^2\rho_{\rm
    crit}^2}{m_\chi^2} \cr
&& \!\!\times \int_0^{z_{\rm up}}\!{\frac{f(z) (1+z)^3}{h(z)}
  \frac{dN(E^\prime)}{dE^\prime} e^{-\tau(z,E^\prime)} dz} ,
\end{eqnarray}
where $H_0 = 70\rm\ km\ s^{-1}\ Mpc^{-1}$ is the Hubble constant,
$\Omega_m$ is the matter density in units of the critical density,
$\rho_{\rm crit}$, and the fraction of matter in dark matter is
$f_{\rm DM} = \Omega_{\rm DM}/(\Omega_{\rm DM}+\Omega_b) \approx
0.833$, where the fraction of critical density of the dark matter we
take is $\Omega_{\rm DM} = 0.237$, and baryon density $\Omega_b =
0.0456$~\cite{Komatsu:2010fb}.  The uncertainties in the halo modeling
are much larger than the errors on the cosmological parameters we
adopt here.  However, much previous work sets $f_{\rm DM} = 1$, which
overestimates the extragalactic annihilation signal by approximately
30\%.  We take a flat universe, with $\Omega_\Lambda$ and $h(z) =
[(1+z)^3 \Omega_m + \Omega_\Lambda]^{1/2}$.  The factor
$e^{-\tau(z,E^\prime)}$ takes into account attenuation of
$\gamma$-rays along cosmological distances, for which we use the
results of Gilmore et al.~\cite{Gilmore:2009zb}.  The factor $f(z)$
accounts for the increase in density squared during halo growth and
the redshift evolution of the halo mass function.  We adopt a fit to
this evolution such that \cite{Ullio:2002pj,Ando:2005hr,Yuksel:2007ac}
\begin{equation}
f(z) = f_0 10^{0.9\left[\exp\left(-0.9 z\right) -1\right] - 0.16 z}.
\label{fando}
\end{equation}
The halo internal density profile sets $f_0$, and for the Einasto
case, we find $f_0 \simeq 3\times 10^4$.  We also include a the boost
factor here of Eq.~(\ref{totalboost}) to the full extragalactic
component, consistent with our adoption for the MW.  The extragalactic
contribution to the diffuse isotropic background is subdominant to the
Galactic contribution, at the level of 10\%-20\% of the total signal
for a given model.  For comparison, shown in
Fig.~\ref{comparison_plot}, our model of the diffuse extragalactic
background is similar to that named ``MS-II Res'' in
Ref.~\cite{Abdo:2010dk}, due to a comparable low mass halo cutoff,
excluding our adopted boost factor.

We redshift the energy distribution of the annihilation photons as $E
= E^\prime/(1+z)$, where $E^\prime$ is the energy of the photons at
the cosmological source.  Eq.~(\ref{fando}) uses a minimal limit of
the halo mass of $10^6\ M_\odot/h$, a conservative lower limit that
only uses cosmological halos known to fit cosmological determinations
of the halo mass functions with no extrapolation~\cite{TinkerMF}.
Note that the limit from the diffuse flux greatly increases with an
extrapolation to extremely low masses such as $10^{-9}\ M_\odot$
employed by H\"utsi et al.~\cite{Huetsi:2009ex}.  With our framework,
we use the observed cosmological diffuse spectrum as the maximum
amount that the annihilation signal can be an a given energy, with
95\% CL limits to the full spectrum for any given $m_\chi$.  For the
95\% CL limit, we sum take the bin-summed $\chi^2 < 1.282$ for this
form of a one-sided upper limit.  These limits are shown in
Fig.~\ref{array_plot}.

\subsection{PAMELA and Fermi Electron/Positron-Spectrum Motivated Models}
\label{pamela}
The PAMELA observation of an increase on the positron fraction at 10
to 100 GeV, in combination with a feature in the shape of the Fermi
$e^+/e^-$ spectrum at $\sim$1 TeV could be consistent with the
production of the high energy $e^+/e^-$ in the products of dark matter
annihilation.  Since annihilation modes to charged particles also
produce photons from bremsstrahlung, the signal is should also be
observed or constrained by the Fermi-LAT isotropic diffuse spectrum.
Channels through the quarks and massive $W^\pm$ and $Z$ bosons are
strongly excluded by a number of observations~\cite{Meade:2009iu}.  We
examine here constraints on two-body charged annihilation modes from
the Fermi-LAT isotropic diffuse background, with 95\% CL exclusions
shown in Fig.~\ref{pamela_plot1}.  The PAMELA 99\% CL preferred region
is in pink, and combined Fermi-$e^+/e^-$ preferred 99\% CL region in
red, from Ref.~\cite{Meade:2009iu}.

We again choose the most conservative cases for the MW halo profile,
as in \S\ref{dmhalo}, with a shallow Einasto profile as that
incorporated in Eq.~(\ref{j_hessgr}), and high mass cut-off of the
halo mass function at $M_{\rm min} = 10^6\ M_\odot$.  The Fermi-LAT
diffuse spectrum excludes a large part of the parameter space for the
PAMELA and Fermi-$e^+/e^-$ 99\% CL regions, and when combined with the
HESS Galactic Ridge constraints, exclude all interpretations of the
PAMELA positron fraction and Fermi-$e^+/e^-$ feature as arising from
dark matter annihilation into two-body standard model particle final
states.

We also consider four-lepton annihilation modes that could occur
through an intermediate force carrying vector or scalar boson, $\phi$,
with mass $m_\phi \lesssim 1\rm\ GeV$.  We take scalar bosons of two
cases, $m_\phi = 0.3\rm\ GeV$ and $m_\phi = 0.8\rm GeV$, into either
$4e$ or $4\mu$ final state modes.  They are shown in
Fig.~\ref{pamela_plot2}.  For the more massive $\phi$ particle,
$m_\phi = 0.8\rm GeV$, the resultant lepton states more energetic and
therefore more constrained by their final state radiation by the
Fermi-LAT diffuse and HESS GR observations.  For either $\phi$ mass,
there exist regions of the parameter space that are not excluded at the
95\% CL level by the gamma-ray data we consider here.

For reference, we also show several other constraints that have been
placed in the literature on these models also using Einasto-type
profiles for the MW halo.  For the two lepton modes, we show the
region excluded by ICS radiation of the original final state radiation
products scattering off of cosmic microwave background photons, as
constrained by Fermi-LAT data from the Galactic poles (triple-dot
dashed), from Ref.~\cite{Cirelli:2009dv}, and originally described in
Ref.~\cite{Cirelli:2009vg}.  We also show the region excluded by
synchrotron emission not observed in Galactic radio observations
(dotted line) from Ref.~\cite{Meade:2009iu}, originally described in
Ref.~\cite{Bertone:2008xr}; for the case of $\chi+ \chi\rightarrow
\mu^++ \mu^-$, we show the exclusion regions from final state
radiation constraints in Fermi-LAT observations of Draco (dashed
line), and ICS radiation constraints from Fermi-LAT observations of
Ursa Minor (dot-dashed line)~\cite{Abdo:2010ex}.  In the four lepton
modes, we show Galactic radio synchrotron
constraints~\cite{Meade:2009iu}.  Also, it should be noted that all
models with $m_\chi \lesssim 1\rm\ TeV$ are firmly excluded by the
lack of a $\gtrsim 20\%$ drop in the Fermi $e^+/e^-$ spectrum below 1
TeV~\cite{Meade:2009iu}.  The ICS constraints from public data of
Fermi-LAT observations towards the Galactic
poles~\cite{Cirelli:2009dv} are comparable to and, in many regions,
stronger than our constraints, though such constraints do depend on
ICS modeling of the signal in these regions.

\section{Conclusions}
\label{conclusions}

The LAT aboard the Fermi Gamma-Ray Space Telescope is opening a new
window to detecting the nature of dark matter or constraining it.  We
have derived conservative constraints on WIMP dark matter annihilation
cross-sections for all standard model channels arising from the
observed isotropic diffuse background by the Fermi-LAT.  We have also
re-examined constraints from the HESS Galactic ridge observation.  In
contrast with previous work, our methods use very conservative models
for the cosmological and halo structure giving rise to the signal,
while still remaining consistent with structure formation in a cold
dark matter framework of a WIMP candidate.  We use such conservative
models in the dark matter halo profile in the Galactic signal
contribution, and conservative lower-limits to the surviving halo mass
function in the Galactic and extragalactic contributions to the
isotropic diffuse background.  For example, we do not consider
isothermal halo profiles which do not arise in any CDM halo formation
calculations.  

Our conservative models are therefore stringent and robust constraints
on any WIMP-like annihilating dark matter particle candidates.  The
constraints are at or near the thermal cross-section for annihilation
to hadronic modes, quarks and gluons.  For low masses, $5{\rm\ GeV}
\lesssim m_\chi \lesssim 10\rm\ GeV$, the constraints exclude the
expected thermal-annihilation scale $\langle \sigma_{\rm A} v\rangle
\approx 3\times 10^{-26}\rm\ cm^3\ s^{-1}$ into hadronic final states.
Such low masses are not {\it a priori} excluded by previous particle
physics or cosmological
constraints~\cite{Choudhury:1999tn,Profumo:2008yg}.  Therefore, such
constraints are approaching an observational lower bound on the dark
matter mass for the thermal-relic WIMP.  The constraints from the
isotropic diffuse background presented here are more stringent than
the constraints from the diffuse background limits of the Fermi-LAT
collaboration when including only the extragalactic
component~\cite{Abdo:2010dk}, and comparable to Fermi observations
towards dwarf galaxies~\cite{Abdo:2010ex}, galaxy
clusters~\cite{Collaboration:2010rg}, and the Galactic
poles~\cite{Cirelli:2009dv}, while being more stringent than full-sky
photon limit methods~\cite{Papucci:2009gd}.  With increased
integration time over the life of the Fermi Gamma-Ray Space Telescope
and further constraints on dark matter structure in local and
extragalactic sources, the isotropic diffuse spectrum or other
Fermi-LAT observations will either reveal the $\gamma$-ray products of
WIMP-like dark matter annihilation, or exclude this class of
candidates.

\acknowledgments

We would like to thank John Beacom, Steve Blanchet, Ilias Cholis,
Marla Geha, Michael Gustafsson, Pat Harding, Ted Jacobson, Andrey
Kravtsov, Julie McEnery, Stefano Profumo, Jenny Siegal-Gaskins, and
Ravi Sheth for useful discussions.  In particular, we thank Julie
McEnery, Michael Gustafsson and the Fermi Collaboration for helping
resolve an error in an earlier version of this work.  KA is partially
supported by NSF Grant PHY-0757966; ZC is supported by NSF Grant
PHY-0801323; CK is supported by grant DOE DE-FG02-96ER50959.

\bibliography{master}
\bibliographystyle{h-physrev5}

\end{document}